\documentclass[preprint2]{aastex}

\newcommand{\kms}{\mbox{km\,s$^{-1}$}}

\slugcomment{Revised version}

\shorttitle{Pal5}
\shortauthors{Odenkirchen et al.}

\begin{document}

\title{The extended tails of Palomar 5: A ten degree arc of globular 
cluster tidal debris}

\author{
Michael Odenkirchen\altaffilmark{1}, 
Eva K. Grebel\altaffilmark{1}, 
Walter Dehnen\altaffilmark{2}, 
Hans-Walter Rix\altaffilmark{1},
Brian Yanny\altaffilmark{3}
Heidi Newberg\altaffilmark{4},
Constance M. Rockosi\altaffilmark{5},
David Martinez-Delgado\altaffilmark{1},
Jon Brinkmann\altaffilmark{6},
Jeffrey R. Pier\altaffilmark{7}
}

\altaffiltext{1}{Max-Planck-Institut f\"ur Astronomie, K\"onigstuhl 17, 
D-69117 Heidelberg, Germany; odenkirchen@mpia-hd.mpg.de}
\altaffiltext{2}{Astrophysikalisches Institut Potsdam, 
An der Sternwarte 16, D-14482 Potsdam, Germany}
\altaffiltext{3}{Fermi Accelerator National Laboratory, P.O. Box 500, 
Batavia, IL 60510}
\altaffiltext{4}{Dept. of Physics, Applied Physics and Astronomy, 
Rensselaer Polytechnic Institute, Troy, NY 12180}
\altaffiltext{5}{Department of Astronomy, University of Washington, 
Box 351580, Seattle, WA 98195-1580}
\altaffiltext{6}{Apache Point Observatory,P.O.\ Box 59, 
2001 Apache Point Road, Sunspot, NM 88349-0059}  
\altaffiltext{7}{US Naval Observatory, Flagstaff Station, P.O. Box 1149, 
Flagstaff, AZ 86002-1149}  

\begin{abstract}
Using wide-field photometric data from the Sloan Digital Sky Survey (SDSS) 
we recently showed that the Galactic globular cluster Palomar 5 is in the 
process of being tidally disrupted. 
Its tidal tails were initially detected in a 2.5 degree wide band along 
the celestial equator. A new analysis of SDSS data for a larger field now 
reveals that the tails of Pal\,5 have a much larger spatial extent and can 
be traced over an arc of 10$^\circ$ on the sky, corresponding to a projected 
length of 4~kpc at the distance of the cluster. 
The tail that trails behind the Galactic motion of the cluster fades into the 
field at an angular distance of $6\fdg 5$ from the cluster center but shows 
a pronounced density maximum between $2^\circ$ and $4^\circ$ from the center. 
The leading tail of length $3\fdg 5$ extends down to the border of the 
available field and thus presumably continues beyond it.   
The projected width of these tails is small and almost constant (FWHM $\sim$ 
120~pc), which implies that they form a dynamically cold and hence long-lived 
structure.  
The number of former cluster stars found in the tails adds up to about 1.2 
times the number of stars in the cluster, i.e.\ the tails are more massive 
than the cluster in its present state. The radial profile of stellar surface 
density in the tails follows approximately a power law $r^{\gamma}$  with 
$-1.5 \le \gamma \le -1.2$. 

The stream of debris from Pal\,5 is significantly curved, which demonstrates 
its acceleration by the Galactic potential. The stream sets tight constraints 
on the geometry of the cluster's Galactic orbit. 
We conclude that the cluster is presently near the apocenter but has 
repeatedly undergone disk crossings in the inner part of the Galaxy leading 
to strong tidal shocks.
Using the spatial offset between the tails and the cluster's orbit we 
estimate the mean drift rate of the tidal debris and thus the mean mass loss 
rate of the cluster. Our results suggest that the observed debris originates 
mostly from mass loss within the last 2 Gyrs.   
The cluster is likely to be destroyed after the next disk crossing, which 
will happen in about 100~Myr. There is strong evidence against the suggestion 
that Pal\,5 might be associated with the Sgr dwarf galaxy. 

\end{abstract}


\section{Introduction}
Globular clusters are the oldest stellar systems commonly found in the Milky 
Way, having typical ages of 12 to 15 Gyr. They thus represent fossil relics 
from the early formation history of the Galaxy. 
However, the globular clusters we see today are probably not representative 
of the system of Galactic globular clusters at early stages. They may in fact 
be the selected 'survivors' of an initially much more abundant population.
Analytic estimates and numerical experiments predict that on time scales of 
Gigayears globular clusters undergo external and internal dynamical evolution, by which they may suffer a permanent loss of members, and eventually dissolve.
 
One of the major factors governing the dynamical evolution of those clusters 
is the Galactic tidal field.
The tidal field has two important effects: (1) It creates drains through 
which stars are carried away from the outer part of the cluster, and  
hereby truncates the bound part of the cluster to a certain spatial limit 
(von Hoerner 1957, King 1962).
(2) It feeds energy into the cluster through so-called tidal shocks, i.e., 
rapid variations of the strength of the external forces which occur during 
crossings of the Galactic disk or close passages of the Galactic bulge 
(Ostriker, Spitzer, \& Chevalier 1972, Kundic \& Ostriker 1995).   
Detailed simulations of globular cluster dynamics for a variety of masses 
and internal structures and different types of orbits in different Galactic 
model potentials have shown that tidal shocks accelerate the dynamical
evolution of globular clusters and enhance their mass loss in such a way 
that 50\% or more of the present-day globulars will be destroyed within the 
next Hubble time (Gnedin \& Ostriker 1997, including disk-shocks; 
Baumgardt \& Makino 2002, using a model without disk). 
Similarly, it has been shown that the present sample of globular clusters 
is most likely the remainder of an initially much more numerous system of 
clusters, many of which are meanwhile dissolved (Murali \& Weinberg 1997, 
Fall \& Zhang 2001). 
In this way, the spatial distribution, kinematics, and mass function of the 
globular cluster system may have changed a lot. 
Also, the shape of the stellar mass function of individual clusters may have 
changed considerably since their formation because of preferential depletion 
of low-mass stars as a result of mass segregation (Baumgardt \& Makino 2002).

Observations suggest that the Milky Way globular clusters are indeed 
spatially truncated by the Galactic tidal field. 
Measurements of the radial surface density profiles or surface brightness 
profiles of globular clusters (e.g., King et al.\ 1968; Trager, King \& 
Djorgovski 1995) showed that many globulars have profiles that decline more 
steeply than a power law and, by extrapolation,  suggest the existence of a 
finite boundary. Their profiles are often well fit by King (1966) models, 
which are of finite size. 
The estimated limiting radii obtained by extrapolation with King models 
were found to correlate with the clusters' galactocentric distances, and 
the way in which they correlate agrees to what would be expected for 
tidal radii in a Galaxy with a flat rotation curve (Djorgovski 1995).
Nevertheless, a firm observational proof for the predicted mass loss and 
dissolution of globular clusters in the Galactic tidal field has been 
missing until recently.

Using color-magnitude selected star counts, Grillmair et al.\ (1995) 
measured the stellar surface densities of a number of globular clusters to 
lower levels and hence larger radii than earlier studies. They then found 
that at very low levels (typically more than four orders of magnitude below 
the central surface density) the observed profiles frequently exceed the 
prediction from King models and extend beyond the limiting radius of these 
models. 
Similar results were obtained by Lehmann \& Scholz (1997), Testa et al.\ 
(2000), Leon et al.\ (2000), and Siegel et al.\ (2001). 
The observed departures from King models are in some cases associated with 
a break in the logarithmic slope of the profile, which resembles the results 
of numerical simulations of globular clusters, where a break in the radial 
surface density profile marks the transition between the bound part and the 
unbound part of the cluster population (e.g., Oh \& Lin 1992, Johnston et 
al.\ 1999a). 
Hence, these observations suggest that many clusters are surrounded by weak 
haloes or tails of unbound stars that might result from tidal stripping. 
On the other hand, the two dimensional surface density distributions obtained 
by Grillmair et al.\ (1995), Testa et al. (2000), and Leon et al.\ (2000) did 
not clearly confirm this suggestion and left doubts about the reality of the 
observed structures because these were in most cases too complex and diffuse 
to be unambiguously identifiable as tails of tidal debris. 
In fact, contamination by galaxy clustering in the background or by variable 
extinction across the field may have lead to spurious detections of such 
tails. The latter was recently demonstrated by Law et al.\ (2003) for the 
low-latitude cluster $\omega$~Cen. 

Among the much bigger and more massive dwarf spheroidal (dSph) satellites 
in the Milky Way halo, there has been growing evidence that at least the 
Sagittarius dSph, which is the nearest of these systems, is subject to 
very substantial mass loss and produces strong tails of tidal debris.  
The stellar stream from this dSph has meanwhile been detected around the 
whole celestial sphere (see Majewski et al.\ 2003 and references therein).   

Turning back to the globular clusters, deep small-field studies with the 
Hubble Space Telescope (HST) revealed that several of the Galactic globulars 
have luminosity functions that are unusually flat or even decreasing towards 
the low-luminosity end (Piotto, Cool, \& King 1997; De~Marchi et al.\ 1999; 
Piotto \& Zoccali 1999; Grillmair \& Smith 2001). 
This deficiency in low-mass stars could be an indication of tidal mass loss 
(when combined with mass segregation in the cluster) and has frequently been 
interpreted in this sense. However this is not by itself a proof of tidal 
mass loss because (1) the observations do not necessarily represent the 
overall luminosity function of the cluster (spatial variations due to 
mass segregation, either dynamical or primordial), and (2) it might be 
possible that intrinsic differences exist between the overall luminosity 
functions of different clusters.       

The evolution of a cluster depends on its internal parameters and its orbit. 
Ostriker \& Gnedin (1997) presented so-called 'vitality diagrams' for 
globular clusters in the parameter space of half-mass radius, mass, and 
Galactocentric distance, showing in which region of this space a cluster 
should lie in order to survive more than 10~Gyr.  Clusters that do not lie 
in this region, and hence are expected to dissolve due to disk- and bulge 
shocks, are those with large half-mass radius and low mass. 
An extreme example for such an object is the sparse cluster Palomar\,5, 
which has a mass of less than $10^4 M_\odot$ and a half mass radius of 
about 20~pc. 
Since Pal\,5 is also one of the clusters that were found to have an 
atypically flat luminosity function (Smith et al.\ 1986, Grillmair \& Smith 
2001) it presents a particularly interesting test case for tidally-induced 
mass loss. 

The commissioning of the Sloan Digital Sky Survey (SDSS; York et al.\ 2000)
provided deep multi-color CCD imaging in $2.5^\circ$ wide stripe across 
Pal\,5. This allowed a wide-field search for cluster stars  
in the surroundings of Pal\,5 (Odenkirchen et al.\ 2001a, hereafter Paper I). 
A previous investigation of Pal\,5 using photographic plates (Leon et al.\ 
2000) had been compromised by severe problems with contamination from 
background galaxies. The SDSS observations, however, enabled excellent 
separation between stars and background galaxies and an efficient selection 
of cluster stars by color and magnitude. We found strong evidence for 
two massive tails of tidal debris emerging from Pal\,5. These tails showed a 
well-defined characteristic shape and were found to contain about half as 
much mass as the cluster. 

The detection of such tails with clearly identifiable structure has two 
important aspects: (1)~It provides conclusive, direct proof for on-going 
tidal mass loss in a globular cluster. (2)~It reveals unique information on 
the orbit of the cluster and opens a very promising way for investigating the 
gravitational potential in the Galactic halo (e.g., Murali \& Dubinski 1999, 
Johnston et al.\ 1999b).  
In the present paper we describe the analysis of further SDSS data for a 
more than five times larger field around Pal\,5, which have become available
since Paper I. 
The goal of this study is to trace the tidal debris of Pal\,5 to larger 
distances from the cluster in order to obtain a more complete census of 
its mass loss and to constrain the basic properties of the distribution of 
the debris such as its shape and density profile.
As we will show, Pal\,5 is the first globular cluster that exhibits 
fully-fledged tidal tails with a total angular extent of $10^\circ$ on the 
sky.

In \S 2 we provide details about the observations and the photometric data 
derived from them. \S 3 describes the methods used to analyse the data. 
In \S 4 we present the resulting surface density distribution of cluster 
stars and describe the basic features of the tidal tails. \S 5 deals with 
the determination of the cluster's local orbit and its extrapolation to a 
global scale. In \S 6 we derive estimates of the mass loss rate and the 
total mass loss of the cluster. The results are discussed and summarized 
in \S 7, and a brief outlook on future work is given in \S 8.

\section{Observations}
The SDSS is a large deep CCD survey designed to cover 10,000 square degrees 
of sky by imaging in five optical passbands, and by spectroscopy. 
The imaging data are obtained in great-circle drift scans using a large 
mosaic camera on a dedicated 2.5m telescope at Apache Point Observatory, 
Arizona.  (For further information on the survey and its technical  
details see York et al. 2000, Gunn et al. 1998, Fukugita et al. 1996, 
Hogg et al.\ 2001, Smith et al.\ 2002, and Pier et al.\ 2002). 

The data that we use in this study stem from the SDSS imaging runs 
745, 752, 1458, 1478, 2190, 2327, 2334, and 2379, carried out between 
March 1999 and June 2001. The various strips of sky scanned in these runs 
yield complete coverage of a $6\fdg5$ to $8^\circ$ wide zone along the 
equator in the right ascension range from $224^\circ$ to $236^\circ$. 
Hereby we have multi-color photometry for Pal\,5 and its surroundings 
in a contiguous, trapezium-shaped field with an area of 87 square degrees. 
The observations reach down to an average magnitude limit of about 
23.0~mag in $i^*$ and 23.5~mag in $r^*$ (approximate limits of 90\% 
incompleteness). 
Photometric and astrometric data reduction and object classification 
were done by the standard SDSS image processing pipeline (see Lupton et 
al.\ 2001 and Pier et al.\ 2002 for different parts of the pipeline). 
The photometry used here is from before the public data release DR1 and 
hence does not precisely match the final SDSS photometric system.\footnote{By 
convention, the magnitudes in the preliminary system are quoted using 
asterisks.}
However, the preliminary photometric calibration of the data is spatially 
uniform to about 3\% (Stoughton et al.\ 2002). The lack of the final 
calibration does not affect our study since we use the photometry in a 
purely empirical and differential way. 

Our investigation is restricted to objects classified as unresolved sources
(thus eliminating a large number of background galaxies that would 
otherwise contaminate the star counts) and uses object  magnitudes 
derived by point-spread function (PSF) fitting. 
The median internal errors of the magnitudes in $g^*$, $r^*$, and $i^*$ 
are 0.015~mag or better for stars brighter than 18.0 mag (in $g^*,r^*,i^*$ 
respectively), rise to values between 0.023 and 0.035~mag at magnitude 20.0, 
and reach the level of 0.10 to 0.17~mag at magnitude 22.0. 
We confirmed these errors by analysing repeated measurements in 
overlapping scans. The median differences between magnitudes 
from independent observations are between $1.0\times$ and $1.2\times$ 
the median internal errors, showing that the quoted median errors 
provide reliable estimates of the photometric accuracy of these
data.

According to the dust maps of Schlegel, Finkbei\-ner \& Davis (1998) 
the southern and south-eastern part of the field is affected by a 
considerable amount of interstellar extinction while in the 
northwestern part the extinction is much lower. More specifically, 
the extinction in the $g$ band varies from 0.15~mag at the northern 
edge to 0.75~mag at the southern edge of the field.
This corresponds to variations in the reddening of the color index 
$g-i$ in the range $0.07 \le E(g-i) \le 0.34$.
To remove these variations from the observed magnitudes we applied
individual extinction corrections derived from the local $E(B-V)$ 
reddening given by Schlegel, Finkbeiner \& Davis. The resulting 
dereddened magnitudes should properly be named $g_0^*$ etc., however, 
for simplicity, we will suppress the index 0 here. 
Since the reddening data from Schlegel et al.\ represent the integrated 
extinction along the entire line of sight, the magnitudes of stars 
that are in front of the bulk of intervening material would in this 
way become overcorrected. However this is unlikely to happen for stars 
belonging to Pal\,5, which are located more than 20 kpc from the sun and
seen far behind the northern part of the Galactic bulge. Color-magnitude 
diagrams for different parts of the field with different amounts of 
extinction show, that there is no sign of overcorrection 
in the blue edge (main-sequence turn-off) of the halo field star population. 
In the case of significant overcorrection this edge would be inclined 
to the blue with increasing brightness, which is not observed. 
   
Due to variations in the observing conditions the completeness of object 
detection at faint magnitudes is somewhat different from run to run. 
This causes artificial inhomogeneities in the stellar surface density 
of the faintest stars. In order to avoid such effects it was necessary 
to cut the sample at $i^* = 21.8$~mag.
At the bright end we chose a threshold of $i^* = 15.0$~mag because 
brighter stars risk to have saturated images and because none of the 
giants in Pal\,5 is brighter than this limit.   
The resulting data set contains about 940,000 point sources.

\section{Photometric object filtering}
\subsection{Unfiltered sample}
The full set of SDSS point sources with magnitudes in the range $15.0 
\le i^* \le 21.8$ has an average surface density of 3.0 stars per arcmin$^2$ 
and a large-scale surface density gradient of 0.13 stars per arcmin$^2$ 
per degree in the direction of decreasing galactic latitude. Figure~1 shows 
a map of the stellar surface density derived by source counts in pixels of 
$3'\times 3'$.  
There are two strong local density enhancements in this field, at positions 
($229\fdg0$,$-0\fdg1$) and (229\fdg6,$+2\fdg1$) in right ascension, 
declination (J2000). 
The first one represents the remote cluster Pal\,5 while the second one 
is due to the much closer and much richer globular M\,5. The latter is not 
relevant for this paper, except that one must avoid this region when 
investigating the properties of the general field around Pal\,5.
The peak surface density of cluster stars in the center of Pal\,5 is 25.2 
arcmin$^{-2}$, i.e., $8.4\times$ the mean density of the surrounding 
field.

Figure~1 also reveals an arc of very weak overdensity extending northeast 
and southwest of Pal\,5. The results that we will present in \S4.1 confirm 
that these are rudimentary traces of extended debris from Pal\,5. 
This is remarkable since it means that weak traces of the cluster's debris 
are visible even without any particular photometric filtering.
However, the surface density of these features is only on the level of 
$1\sigma$ (rms of surface density in individual pixels) above the local 
mean surface density of the field. 

In order to get detailed information on the distribution of stars from 
Pal\,5 one needs to enhance the contrast between the cluster and 
the field, in particular the dominating Galactic foreground. 
To first order, this can be achieved through simple cuts in color and 
magnitude. A more efficient variant of this method is to use an appropriately 
shaped polygonal mask in color-magnitude (hereafter c-m) space. This approach 
was taken, e.g., by Grillmair et al.\ (1995), Leon et al. (2000), and in 
Paper~I in the context of globular clusters, and also by 
Majewski et al.\ (2000) and Palma et al.\ (2003) in the context of dwarf 
spheroidal (dSph) galaxies.   
However, sharply defined cuts or windows in c-m space do still not provide 
the optimal method to map out the spatial distribution of the stellar 
population of Pal\,5 because each star is treated as either a definite 
member or a definite non-member of the cluster population. This does not 
completely exploit the available information. 
Since the photometry actually allows us to derive smoothly varying membership 
probabilities as a function of color and magnitude one can optimize the 
object filtering by using these probabilities.

\subsection{Optimal contrast filtering}
  
A straight-forward way to make comprehensive use of the photometric 
information is to construct empirical c-m density distribution templates
$f_C(m,c)$, $f_F(m,c)$ for the cluster and the field ($m$ and $c$ denoting 
magnitude and color index), and 
to use these to determine the surface density $\Sigma_C$ of the cluster 
population for each position in the field by a weighted least-squares 
adjustment. This kind of approach was, e.g., described by Kuhn, Smith, \& 
Hawley (1996) in a study of the Carina dSph, and recently discussed in more 
detail by Rockosi et al.\ (2002), who used it for an analysis of a smaller 
SDSS data set on Pal\,5. 
The adjustment is done such that the weighted sum of field stars plus cluster 
stars yields the best approximation of the observed total c-m distribution. 
In the present study we have applied this method of optimal data filtering in 
the following way:

Using the magnitudes $g^*$, $r^*$, and $i^*$ (which provide higher accuracies 
than $u^*$ and $z^*$), we first defined orthogonal color indices $c_1$ and 
$c_2$ as in Paper~I:

\begin{mathletters}
\begin{eqnarray}
c_1 & = & \ 0.907 (g^* - r^*) + 0.421 (r^* - i^*)\ \ \\
c_2 & = & -0.421 (g^* - r^*) + 0.907 (r^* - i^*)\ \
\end{eqnarray}
\end{mathletters}
 
The choice of the indices is such that the main axis of the almost 
one-dimensional locus of Pal\,5 stars in the $(g-r)$ versus $(r-i)$ 
color-color plane lies along the $c_1$-axis while the $c_2$ axis is 
perpendicular to it. 
We then preselected the sample in $c_2$ by discarding all objects with 
$|c_2| > 2\sigma_{c_2}(i^*)$, where $\sigma_{c_2}(i^*)$ is the rms 
dispersion in $c_2$ for stars of magnitude $i^*$ in Pal\,5. 
Stars with these $c_2$ colors are unlikely to be from the cluster. 
We also preselected in $c_1$ by restricting the sample to the range 
$0.0 \le c_1 \le 1.0$ because one expects very few stars of the cluster 
population outside this range.

Next we constructed c-m density diagrams (Hess diagrams) for the cluster 
and the field by sampling the stars that lie within $12'$ from the center 
of Pal\,5 (cluster diagram) and those that are more than $1^\circ$ away from 
the location of the stream and outside M\,5 (field diagram) on a grid in 
the plane of $c_1$ and $i^*$. 
Bins of 0.01\,mag$\times$\,0.05\,mag were used and the counts were 
smoothed with a parabolic kernel of radius 3 pixels.\footnote{In 
order to treat the bins at the borders of the c-m domain in the same way 
as in the interior the grid counts were actually extended somewhat beyond 
the c-m limits specified above.}
The cluster c-m distribution was corrected for the presence of field stars 
in the $12'$ circle around the cluster center by subtracting the field c-m 
distribution in appropriate proportion. 
The resulting diagrams of the normalized c-m densities $f_C$, $f_F$ of 
cluster stars and field stars are shown in Figures~2a and 2b. 
The cluster members are concentrated along well-defined branches (giant 
branch, horizontal branch, subgiant branch and main sequence) while the 
field star distribution is more diffuse, showing local maxima along 
$c_1 \approx 0.4$, which can be attributed to the turn-off region of thick 
disk and halo.\footnote{Note that a substantial fraction of field stars 
actually lies at $c_1 > 1.0$ and was already eliminated by the 
preselection in $c_1$ and $c_2$}
By comparing the field star c-m distribution in the region northwest of 
the cluster to that in the region southeast of the cluster it appears that 
$f_F$ is not strictly constant over the field. However, the differences are 
not dramatic because the deviations from the mean distribution are mostly 
below 10\%.  
Since a more local estimate of $f_F$ can only be obtained at the cost of 
higher noise or lower c-m resolution we preferred to neglect the spatial 
variations and to work with the mean distribution shown in Figure 2b.

In order to derive the surface density distribution of cluster stars on the 
sky one needs a mathematical model that provides a link to the observed 
distributions. The general ansatz for the stellar density in the hyperspace 
spanned by the celestial sphere and the c-m plane is a sum of densities 
$S_C$ and $S_F$ for the cluster and the field (i.e., non-cluster stars)

\begin{mathletters}
\begin{eqnarray}
S(\alpha,\delta,m,c) 
& = & S_C + S_F \\
S_C & = & \Sigma_C(\alpha,\delta)\Phi_C(m,c) \\
S_F & = & \Sigma_F(\alpha,\delta)\Phi_F(\alpha,\delta,m,c)
\end{eqnarray}
\end{mathletters}

where each component can be split up into a product of a surface density 
$\Sigma$ on the sphere and a position-dependent normalized c-m density $\Phi$.
($\alpha,\delta$ denote coordinates on the celestial sphere, $m$ and $c$ 
denote magnitude and color index.)
For the cluster component as a sample of stars of common origin we assume
that (1) it is everywhere composed of the same mix of stellar types and that 
(2) all stars are at practically the same distance from the observer. This 
implies that $\Phi_C$ does in fact not depend on position ($\alpha,\delta$)
(as in equation 2b). 
In contrast to this the field component is an inhomogeneous sample,
i.e., its composition by stars of different types and its density 
distribution along the line of sight are spatially variable, so that $\Phi_F$ 
must in principle vary with position on the sky (equation 2c).  

Let $j$ be an index labelling the pixels of a grid in the c-m plane and 
$k$ be an index labelling the positions $(\alpha_k,\delta_k)$ of 
a grid on the sky, then the number $\nu(k,j)$ of stars lying in a solid angle 
$\Omega_k$ centered on $(\alpha_k,\delta_k)$ and with magnitude and color
falling on the pixel $j$ of area $P_j$ is obtained by integrating equation
(2a) over $\Omega_k$ and $P_j$, i.e.

\begin{eqnarray}
\nu(k,j) &=& \nu_C(k) f_C(j) + \nu_F(k) f_F(k,j) \quad\\
\nonumber\\
\mathrm{where} \nonumber\\
\nu_{\,C,F}(k)&=&\int_{\Omega_k} S_{\,C,F}\ d\Omega \nonumber\\
f_C(j)&=&\int_{P_j} \Phi_C\ dm\,dc \nonumber\\
f_F(k,j)&=&\frac{1}{\nu_F(k)}\int_{\Omega_k} \int_{P_j} 
\Phi_F\ dm\,dc\,d\Omega \nonumber
\end{eqnarray}

Although $f_F$ does in principle depend on position index k, this dependence 
is in our case not important because we can assume that substantial changes 
in the characteristics of the field population occur only on larger scales 
and that hence (as the observations show) $f_F$ is approximately constant 
within the chosen field. 
\footnote{This assumption is of course violated at the location of the 
foreground cluster M\,5.}  
The distributions $f_C$ and $f_F$ in the model of equation (3) can thus be 
represented by the above normalized average c-m distributions that have been 
drawn from the observations. 
$\nu_C(k)$ and $\nu_F(k)$ are the numbers of cluster stars and field 
stars in $\Omega_k$, the former of them being the target of our analysis.   
Apart from observational noise (and apart from small deviations due to 
spatial variations in the c-m distribution of the field stars, which the 
model neglects), the left hand side of equation (3) should correspond to 
the observed star counts $n(k,j)$ in $\Omega_k \times P_j$. Thus one can 
plug in $n(k,j)$ for the expected number $\nu(k,j)$ in equation (3).   
Equation (3) then does not have an exact solution.  
However, we can determine a least-squares solution for $\nu_C(k)$ by 
demanding that the square sum of the noise-weighted deviations between 
the observed number $n(k,m)$ and the expected number $\nu(k,m)$ given 
by the right hand side of equation (3), summed up over the c-m grid, 
is minimal.
Since the contribution of the cluster population to the total counts 
is small (outside the cluster) we assume the noise to be dominated by 
the field stars, i.e., we expect $\sigma_n^2(k,m) = \nu_F(k) f_F(m)$. 
The sum of weighted squares to be minimized thus is:      

\begin{eqnarray}
\chi^2(k) = \hspace*{5.5cm}\\
\quad \sum_j \frac{\left(n(k,j) - \nu_C(k) f_C (j) - 
\nu_F(k) f_F(j)\right)^2}{\nu_F(k) f_F(j)} \nonumber
\end{eqnarray}

It is straightforward to calculate (via $d\chi^2/d\nu_C = 0$) that the 
least-squares solution for $\nu_C(k)$, which we call $n_C(k)$, and its 
variance $\sigma^2_{n_C}(k)$ are:

\begin{mathletters}
\begin{eqnarray}
n_C(k) = \frac{\displaystyle \left(\sum_j n(k,j)\frac{f_C(j)}{f_F(j)}
\right) - n_F(k)}{\displaystyle \sum_j\frac{f_C^2(j)}{f_F(j)}}\\
\sigma^2_{n_C}(k) = \quad n_F(k) \bigg/ \sum_j 
\frac{f^2_C(j)}{f_F(j)} \qquad
\end{eqnarray}
\end{mathletters}

In principle, one could determine both $n_C$ and $n_F$ (i.e., the best 
estimate for $\nu_F$) in this way by minimizing the $\chi^2$ of equation (4). 
However, for $\nu_F$ we already know (or can safely assume) that it must be 
a smoothly varying function of position that can be described by a simple 
(polynomial) model. 
Thus we preferred to use this constraint to determine $n_F$ externally 
(for practical details see \S 4.1) and not in a simultaneous least-squares 
adjustment with $\nu_C$. This makes the solution for $\nu_C$ more robust. 

Equation (5a) allows the following interpretation: One finds the 
least-squares solution $n_C(k)$ by weighting each star in the solid 
angle $\Omega_k$ by the quotient $w(j)=f_C(j)/f_F(j)$ according to its 
position in the c-m plane, summing up the weights of all stars, and dividing 
this sum by the factor $a = \sum_j (f_C^2/f_F)$.
This yields the estimated number of cluster stars $n_C$ plus a term $n_F/a$, 
i.e., the estimated number of field stars attenuated by $a$. 
By subtracting this residual field star contribution one obtains $n_C$.  
Equation (5b) shows that the variance of $n_C$ is reduced to $1/a$ times the 
variance of the field star counts. In other words, the noise in the 
determination of the surface density of cluster stars decreases by the factor 
$\sqrt{a}$.   
The weight function $w=f_C/f_F$ is shown by the contour plot in 
Figure~2c. We obtain an attenuation factor $a=5.1$ and hence a noise 
reduction of $\sqrt{a} = 2.3$ with respect to the unfiltered, but 
preselected sample. In total, i.e., in comparison with the full sample, 
the noise level is reduced by a factor of 4.3.\\

\section{The tidal stream}
\subsection{Surface density map}

We constructed a map of the stellar surface density of Pal\,5 stars by 
applying the above method of least squares estimation on a grid with pixels 
of $3'\times 3'$ in the plane of the sky. The residual contribution from 
field stars was determined by fitting a bi-linear background model to the 
weighted counts in those pixels that are at least $1^\circ$ away from the 
cluster and the tails, and also away from M\,5. 
After pre-selection and weighting, the mean surface density of the field 
stars is 0.16 arcmin$^{-2}$ and the surface density gradient across the field 
is about $5\times 10^{-3}$ arcmin$^{-2}$ per degree. 
By subtracting the best-fit bi-linear background model the density 
distribution in the field becomes essentially flat. 
For further reduction of the noise the counts were smoothed by weighted 
averaging with a parabolic kernel of radius 4 pixels. In regions with 
surface density above the 5$\sigma$ level we used a kernel with a smaller 
radius to preserve higher resolution. Figure~3 presents the resulting 
surface density map as a grey scale and contour plot in equatorial celestial 
coordinates (i.e., right ascension and declination). 

This map shows a striking coherent structure that is spatially connected 
to the main body of Pal\,5 and has stellar surface densities 
varying from $1.5 \sigma$ up to $5\sigma$ ($\sim$ 0.12 arcmin$^{-2}$) 
and higher. The geometry relative to the cluster clearly identifies 
this structure as debris from Pal\,5. The debris forms two long narrow 
tails on opposite sides of the cluster and extends over an arc of about 
10$^\circ$. This corresponds to a projected length of 4~kpc at the distance 
of the cluster. 
The tails have a width of about $0\fdg 7$ in projection on the plane 
of the sky. Apart from small scale variations the width of the tails 
does not systematically change with angular distance from the cluster. 
The northern tail, which - as will be shown later in \S5.2 - is trailing 
behind the cluster, is traced out to an angular distance of at least 
$5\fdg 8$ from the center of the cluster, but possibly out to $6\fdg 5$, 
and appears slightly curved. 
The maximum surface density of stars in this tail is about 0.2 arcmin$^{-2}$. 
It occurs at angular distances between $2\fdg 2$ and $3\fdg 7$ 
from the cluster center and was hence not covered by the initial detection 
of the tails in Paper I.

The southern tail, which is the one that leads the motion of the cluster 
(see \S5.2), 
is seen over $3.6^\circ$ and reaches down to the edge of the currently 
available field. This suggests that the tail continues beyond this limit, 
as one would indeed expect when assuming approximate point symmetry in the 
distribution of debris with respect to the cluster. 
The southern tail exhibits density maxima at angular distances of 
$1\fdg 6$ and $3\fdg 5$ from the cluster center, which are, however, less 
pronounced than in the northern (trailing) tail. 
The transition between the cluster and the tails is not straight. Instead
we see a characteristic S-shape in the distribution of stars, which closely 
resembles the structures seen in simulations of disrupting globular clusters
and satellites (e.g., Combes et al.\ 1999, Johnston et al.\ 2002). 
This S-shape feature clearly indicates that the stars are stripped from 
the outer part of the cluster by the Galactic tidal field dragging them away 
in the direction towards the Galactic center and anticenter. 

Besides the two tails of Pal\,5 and a spurious patch of high stellar density 
left over from the cluster M\,5, the map shows a number of small isolated 
patches with densities on the level of $2\sigma$ above zero, which are 
dispersed over the field. These are most likely not traces of the population 
of Pal\,5 but the result of random fluctuations in the distribution of field
stars. To test this we generated random fields with the same mean density as 
the observed residual field star density and sampled these artificial fields 
in exactly the same way as done with the observations. This Monte Carlo 
experiment showed that random fields yield $2\sigma$-patches of the same size 
and with very similar number densities as in Figure~3.
As another statistical test we resampled the observations on a grid of 
non-overlapping pixels with a size of $9' \times 9'$ and determined the
frequency distribution of pixel counts in those regions that lie outside 
the clusters M\,5 and Pal\,5 and the tails of Pal\,5. This distribution 
closely agrees with the expected Poisson distribution for a random field 
of the given mean density (see similar tests of fluctuations described in 
Odenkirchen et al.\ 2001b). Both tests reveal that the isolated patches  
in the map of Figure~3 provide no evidence for further significant local 
overdensities.

In order to estimate the fraction of stars in the tails and in the cluster, 
we integrated the surface density of Pal\,5 stars in a $42'$ wide band 
covering the tails (2.3 times the FWHM of the tails, see \S 6.1) and in a 
circle of radius $12'$ around the center of the cluster. A somewhat smaller 
radius than the cluster's limiting (or tidal) radius of $16'$ (Odenkirchen 
et al.\ 2002, hereafter Paper II) was used because the bound and unbound 
part of the cluster overlap in projection on the sky and cannot be strictly 
separated. 
We find that the number of stars in the tails is about 1650 while the number 
of stars in the cluster is about 1350. This yields a number ratio 
between tails and cluster of $\beta = 1.22$. The number of stars seen 
in the southern (leading) tail is about half the number found in the 
northern (trailing) tail (i.e., there are about 1100 stars in the northern 
and about 550 stars in the southern tail). 
To check these number ratios we also took an alternative approach and 
performed integrated number counts on a variety of samples defined by 
different choices of cut-off lines in the c-m plane. 
Hereby we obtained values for the ratio between the number of stars in the 
tails and in the cluster in the range from 1.18 to 1.31. 
It thus appears that $\beta = 1.25\pm 0.06$ is a robust estimate for the 
observed field. Since the tails may easily extend beyond the area 
currently covered by the SDSS, the 'true' ratio is likely to be higher. 
In any case, we can safely conclude that the tails contain more stellar 
mass than the cluster.

\subsection{Density profile along the tail}
Since the debris forms a long and relatively thin structure,
it makes sense to treat it as a one-dimensional object and to characterize 
it by its distribution of linear density. 
To determine this linear density we modelled the central line of each tidal 
tail by a sequence of short straight-line segments (fitting by eye) and 
projected all stars within $0\fdg 35$ from the central line perpendicularly 
onto it. 
We then performed weighted star counts as described in equation (4a) in bins 
of the arc length parameter $\lambda$ along the central line, using a bin 
size of $0\fdg 25$.
The field star contribution was determined using the bilinear background 
model from Section 4.1, and was subtracted from the counts.
We thus obtained the density distributions shown in Figure~4 (statistical 
uncertainty of the number counts indicated by error bars).

The density curve for the northern (trailing) tail (Fig.4a) shows three 
pronounced maxima, which correspond to extended density clumps around 
RA $230\fdg 9$, $231\fdg 8$, and $233\fdg 3$ in the map of Figure~3. 
The linear density at these maxima is about two times as high as it is 
on average. Apart from those local variations there is a general decline 
in the density with increasing $\lambda$.  
The mean density level decreases by roughly a factor of 3 when comparing 
$\lambda\sim 0\fdg 5$ to $\lambda\sim 6^\circ$. 

In order to judge the significance of the observed variations we approximate 
the general trend in the data by fitting a straight line to the innermost 
five and the outermost five data points (dashed line in Fig.4a). Comparing 
the data with this line shows that the smaller amplitude variations in the 
counts lie within the error bars and hence are likely of statistical nature 
whereas the strong maxima at the above given locations exceed the 
straight-line model by about 3 times the error bar. 
Therefore, these maxima are statistically significant and present real 
substructure in the tail. Two of the clumps may in fact be part of one broad 
density enhancement because their separation by only one bin of lower density 
could be the result of statistical fluctuations.  

Along the southern (leading) tail (Fig.4b) the linear density is generally 
lower than in the corresponding part of the northern (trailing) tail. Again, 
there are local variations which reflect the presence of density clumps 
in Figure~3. However, these variations have lower amplitude than those 
occuring in the northern tail, and the deviations from the general trend of 
the data (fit of straight line to entire set of data points) are thus not 
highly significant.
In particular, the southern (leading) tail shows no obvious counterpart to 
the broad density enhancement in the northern (trailing) tail. 
Since the data for the southern (leading) tail cover a smaller range in 
$\lambda$ there is less information on the large-scale trend of the 
density. With the exception of the outermost bin the data points seem to 
suggest a weak outward decrease. 
On the other hand, taking into account the error bars the counts are 
also consistent with the assumption of a constant mean density level.
In any case, the density curve leaves no doubt that the southern tail 
must continue beyond the border of the field.
The steep rise of the counts in the outermost bin, be it a statistical 
fluctuation or due to a real clump, shows that the mean density does not 
drop to zero at this point.
Whether or not the linear density at higher $\lambda$ declines in a similar 
way as seen in the outer part of the northern (trailing) tail is an  
interesting question that can presently not be answered.

\subsection{Radial profile of the surface density} 
Another way to describe the tidal debris is by determining the radial 
profile of the surface density, i.e., the azimuthally averaged surface 
density as a function of distance from the cluster center. This description 
disregards the fact that tidal tails are not a circularly symmetric 
structure, but has the advantage to provide a uniform view of both the 
cluster and the debris. 
Therefore, observational studies of globular clusters and local dwarf 
galaxies are often judging the existence of tidal debris in this way, and 
results of theoretical studies are also frequently presented in this form 
(e.g., Johnston et al.\ 1999a, Johnston et al.\ 2002). 

We derived the radial profile of the cluster and the two tails through 
weighted number counts in sectors of concentric rings. Out to $r = 15'$ 
we divided each ring into its northern and southern half. At larger radii 
we used progressively narrower sectors to bracket the tails and to minimize 
the influence of the field, but referred the (background corrected) counts 
to the full area of the corresponding half ring. This yields the profiles 
shown in Figure~5. For comparison we also show an analogous profile obtained 
in two cones away from the tidal tails, i.e., at position angles 
$100^\circ\pm 35^\circ$ and $280^\circ\pm 35^\circ$. 

It is clearly visible that the tidal debris is distinguished from the cluster 
by a characteristic break in the slope of the logarithmically plotted radial 
profile.
Outside the cluster's core region, i.e., at radii $r > 3'$, the surface 
density first decreases steeply as a power law $r^\gamma$ with exponent
$\gamma = -3$. 
Between $15'$ and $20'$ there is a transition region where the profile 
becomes less steep, and from $20'$ outwards the decline of the density is 
similar to a power law with an exponent in the range $-1.5 < \gamma < -1.2$. 
The comparison profile, which has been measured perpendicularly to the tails 
and should thus not be affected by tidal debris, shows the same $r^{-3}$ 
power law decline between $3'$ and $10'$ but falls off more steeply at 
$r > 10'$. 
This shows that perpendicular to the tails the cluster has a well-defined 
radial limit. A fit of a King profile to these counts suggests a limiting 
(or tidal) radius of approximately $16'$ (see Paper II). This is near to 
the radius where the overall radial profile shows the break.
By comparing the different radial profiles the tidal perturbation of the 
cluster is noticable from about $r = 12'$ outwards. 

To determine the power law exponent for the outer part of the radial profile 
we made a weighted least-squares fit to the data points at $r \ge 20'$. For 
the southern (leading) tail this fit yields $\gamma = -1.25 \pm 0.06$. 
For the northern (trailing) tail the use of all data points results in a poor 
fit with $\gamma = -1.36$. 
When leaving out the three most discrepant data points, which describe the 
strong local density maximum in the range $140' < r < 220'$, we obtain an 
acceptable fit and $\gamma = -1.46 \pm 0.06$. The overall decline of the 
radial surface density profile of the northern (trailing) tail is thus 
somewhat steeper than for the southern (leading) tail. For both tails we 
find power law exponents $\gamma < -1$, which means that the decline is 
steeper than it would be for a stream of constant linear density 
(having a radial profile $\propto r^{-1}$ because the area of the 
averaging annuli increases proportional to $r$). This confirms that the 
linear density of the stream is decreasing with angular distance 
from the cluster as stated in \S 4.2. On the other hand, it also reveals
that the decrease in linear density is distinctly less steep than $1/r$ 
because we find $\gamma \ge -1.5$.  

\subsection{Distances} 
It is important to recall that our mapping of the tidal debris is built 
on the assumption that the debris is located at the same heliocentric 
distance as the cluster (at least within the limits of the photometric 
accuracy and the natural photometric dispersions). For the immediate 
vicinity of the cluster this necessarily holds true. With increasing 
angular distance from the cluster the heliocentric distances might however 
increasingly deviate, depending on how much the tidal stream is inclined 
against the plane of the sky. 
If, for example, this inclination were $\ge 50^\circ$ the distances 
should differ by $\ge 10\%$ over an angle of $5^\circ$, resulting in 
shifts of $\pm 0.2$\,mag or more in apparent magnitude.
One might suspect that shifts of this size, if true, could affect our 
measurements of the stellar surface density along the tails. 
On the other hand if such shifts in apparent magnitude were detectable, 
this would also provide interesting constraints on the extent 
of the tidal debris and the cluster's orbit in the third dimension .

Unfortunately, the stars that we have access to in the tails are not well 
suited as precise distance indicators. In order to measure small distance 
effects we would ideally need stars with characteristic luminosities such 
as horizontal branch (HB) stars. These are not very numerous, even in the  
main body of the cluster ($\approx$ 30 HB candidates within $12'$ from the 
center including variables), and occur mostly on the red side of the HB. 
In the tails an occasional red HB star from Pal\,5 would (in the absence 
of kinematic information) be indistinguishable from Galactic field stars. 
The subgiant branch is also not sufficiently well populated to allow such 
cluster members to be recognized on a purely statistical basis. Therefore, 
one has to rely on stars near and below the main-sequence turn-off, whose 
luminosities cover a wider range. Even for stars of this type one needs to 
integrate over a substantial part of the tails in order to be able to 
identify their location in the c-m plane. Therefore, distance variations 
can only be investigated at low angular resolution. 

In Figure~6 we present Hess diagrams for the outer parts of the two tails, 
obtained by sampling stars in two $18'$ wide bands ($\approx$ the FWHM of 
the tails, see \S 6.1) along the ridge lines of the tails. Panel (a) of 
this figure shows the integrated c-m distribution in the northern (trailing) 
tail between 3\fdg5 and 5\fdg6 from the center of Pal\,5, 
while panel (b) shows the same for the southern (leading) tail from 1\fdg 5 
to its outer end. The two samples, which have almost the same size, are thus 
spatially separated by an angle of at least $5^\circ$. 
For comparison with the cluster, the ridge line of the cluster's
c-m distribution as derived from Figure~2a is overplotted (middle dot-dashed 
line) and repeated with magnitude offsets of $-0.2$ and +0.2 mag (upper and 
lower dot-dashed line, respectively), corresponding to a 10\% smaller or 
larger mean distance of the stars. The location of the density maxima in 
these diagrams reveal that the outer part of the northern tail is centered 
on the same distance as the cluster while the outer part of the southern 
tail appears to be about 0.1 mag brighter. Hence its mean distance may be
about 5\% smaller. The fact that the c-m distribution of the 
southern tail sample extends to brighter magnitudes also appears to be 
influenced by field stars with $c_1 \approx 0.35$, which are seen to be more 
abundant than in the northern sample and spread into the locus of the cluster 
members. 
Thus the mean distance of the southern (leading) tail sample is probably not 
smaller than that of the cluster by as much as 10\% (i.e., $-0.2$~mag) or 
more. 
Accepting a relative difference of 5\% between the mean distances of the 
northern (trailing) and the southern (leading) tail sample and considering 
that the mean angular separation between those two samples is 7\fdg1, the 
inclination of the tidal stream against the plane of the sky may be of the 
order of $22^\circ$. 
Since the data shown in Figure~6 do not support a difference $\ge 10\%$ in 
the mean distances of the two samples an inclination of $\ge 38^\circ$
can be excluded.  
 
To determine how variations in heliocentric distance along the southern 
(leading) tail might influence the determination of the stellar surface 
density in the stream we shifted the color-magnitude distribution $f_C$ of 
the cluster by $-$0.1~mag and $-$0.2~mag, recomputed the weight function, and 
rederived the least-squares solution for the surface density. 
Figure~7 shows the resulting linear density profiles along the southern 
(leading) tail. One can see that the above magnitude shifts in the cluster 
template lead to slightly lower densities in most of the bins. The general 
trend of the data with arc length along the tail as determined by the 
best-fit straight line (dashed lines in Fig.7) does not change significantly. 
Only in the outermost bin (at $\lambda \sim 3\fdg6$) magnitude shifts of 
$-0.1$ and $-0.2$~mag produce an increase in the number density of stars 
such that the measured density exceeds the general trend by two times the 
statistical error. 
The general conclusion from this experiment thus is that despite a possible 
decrease of the distance along the southern (leading) tail of up to 10\% (out 
to the tip of the tail) the assumption of constant distance as used in the 
previous sections does not cause a significant underestimation of the stellar 
surface density in the outer part of this tail.

\subsection{Luminosity functions}
Another assumption in the filtering method described in \S 3.2 is that 
the tidal debris has the same luminosity function and c-m distribution 
as the cluster. This is not necessarily the case because there could be 
mass segregation effects (see \S 7.4). 
However, using star counts in a narrowly confined band containing the tails 
it can a posteriori be shown that down to our magnitude limit of $i^*=21.8$, 
which comprises only a small range in stellar mass, the assumption holds 
true. In Figure~8 we present the luminosity function of the stellar 
population in the tidal tails and compare it to the luminosity function of 
the stars in the cluster itself. These luminosity functions were obtained 
by restricting the star counts to an appropriate window around the loci 
of the giant branch, subgiant branch, and upper main sequence of Pal\,5
in the c-m plane. The luminosity function of the cluster was derived by
counting stars within $r \le 6'$ from the cluster center. The luminosity 
function of the tails was obtained by counts within $\pm0\fdg25$ angular 
distance from the central line through the tails (see \S 4.2). 
Possible variations in the line-of-sight distances of the stars along the 
tails were neglected since their effect in apparent magnitude is small.      
The statistical contamination by intervening field stars was determined with 
counts in neighboring zones and subtracted after proper scaling with the 
respective areas.  Finally, the luminosity function of the tails was 
renormalized in order to bring it to the same level as the cluster's 
luminosity function in the magnitude bins centered on $i^* = 19.0$ and 
$i^*=19.5$ (renormalization factor of 100).       
Figure~8 shows that the two luminosity functions are almost identical from 
$i^*=19.0$ down to the faintest bin. At magnitudes brighter than 18.5 
the number of stars in the tails is too small to decide whether or not the 
surface density is higher than in the field. However, within the statistical 
errors the counts agree with the number of giants in the cluster, which is 
also quite small. Anyway, stars of this brightness are unimportant in the 
filtering process. The agreement between the two luminosity functions proves 
that the filtering method is based on firm grounds.

\section{Implications on the orbit of Pal\,5}
Numerical simulations of globular clusters in external potentials demonstrate 
that stars tidally stripped off from such systems remain closely aligned with 
the orbit of the cluster (e.g., Combes et al.\ 1999).
This happens because the stars have only small differential velocities and 
small spatial offsets when they become unbound from the cluster.
N-body simulations for dSph satellites have shown that even debris from
such more massive systems may remain on approximately the same orbit as the 
parent object over long time-scales (e.g., Johnston et al.\ 1996, 
Zhao et al.\ 1999). 
The stream of debris from Pal\,5 thus provides a unique tool for tracing 
the orbital path of this cluster and subsequently also its orbital kinematics.
This offers an exceptional opportunity to probe the Milky Way's potential 
with observations of a Galactic orbit. So far, only the Sgr dSph galaxy with 
its global tidal stream has allowed to constrain the Galactic potential in 
a similar way (see Ibata et al. 2001, Majewski et al.\ 2003). 
Classically, determinations of the potential of the Galactic halo 
have been based on statistical investigations of the bulk properties (spatial 
distribution and average velocities) of object samples, either halo stars, 
globular clusters, or dSph galaxies (e.g., Zaritsky et al. 1989, Kulessa \& 
Lynden-Bell 1992, Wilkinson \& Evans 1999). The availability of traces of 
individual orbits as for Pal\,5 and the Sgr dSph is clearly an advantage 
over the statistical approach because this allows to obtain information on 
the potential without assumptions on, e.g., steady state or velocity 
anisotropy and without dependence on proper motion measures, which are 
often unreliable.

\subsection{Isochrone approximation} In order to make the connection between 
the cluster's Galactic orbit and its tidal tails more transparent we outline 
a simple analytic approach. 
Herein we use a spherical logarithmic halo as a model of the Galactic 
potential and describe the orbits of the cluster and the debris stars with 
the so-called isochrone approximation (Dehnen 1999). In contrast to 
classical epicycle theory, whose application is limited to almost circular 
orbits, this method allows an accurate approximation of substantially 
eccentric orbits. A detailed description for the special case of a 
logarithmic potential is given in Appendix A. 
The key point is that using appropriate transformations of the radial 
coordinate $R$ (galactocentric distance) and the time parameter $t$, the 
radial motion can be described by a harmonic oscillation whose period $T_R$ 
is proportional to 
the radius $R_E$ of a circular orbit of equal energy. Furthermore, the 
eccentricity $e$ of the orbit is a function of the quotient $L/R_E$, where 
$L$ is the angular momentum.  

Let us approximate the orbit of the cluster in this way. If at an 
arbitrary point on the orbit (say at $t = t_0$) we shift one star from 
the cluster radially from $R$ to $R' = \alpha R$ ($\alpha > 0$), and 
release it as an independent test particle having the same instantaneous 
space velocity vector as the cluster, then it follows from Eqs.\ (A13) 
to (A19) that the 
eccentricity of its orbit does not change and that the new orbital path of 
this particle is simply a scaled copy of the orbital path of the cluster, 
i.e., $R'(\varphi) = \alpha R(\varphi)$, $\varphi$ being the azimuth or 
phase angle. 
The motion along this path is characterized by a scaling relation 

\begin{eqnarray}
t'(\varphi_2) - t'(\varphi_1) = \alpha (t(\varphi_2) - t(\varphi_1)) 
\end{eqnarray}
between the time parameter $t$ for the cluster and $t'$ for the shifted 
particle, or by the equivalent relation $T' = \alpha T$ for the periods 
of the radial oscillation (e.g., the time intervals between successive 
apocenter or pericenter passages). A star shifted outward ($\alpha > 1$) 
thus has a proportionally longer period and trails behind the cluster 
while a star shifted inward has a proportionally shorter period and leads 
the cluster.  

We now apply this model to the tidal debris of a globular cluster: 
Stars leave the cluster by passing near the inner or the outer Lagrange 
point on the line connecting the cluster to the Galactic center, i.e., 
those points where a force balance between the internal field of the cluster 
and the external tidal field exists.  
They are likely to pass these points with small relative velocity because 
the internal velocity dispersion in the cluster is low, in particular in 
low-mass clusters like Pal\,5 ($\sigma_{los} < 0.7$\,\kms, Paper II). 
Subsequently, 
the debris is decoupled from the cluster and behaves like a swarm of test 
particles that are radially offset from the cluster and released with almost 
the same galactocentric velocity vector as the cluster.     
In the framework of the above model and the ideal case of zero velocity 
dispersion this means that the cluster and its debris are on confocal
orbits that are equal up to radial scaling but have different angular 
velocities and thus exhibit azimuthal shear.

If the separation $\delta\varphi$ between the azimuth angles of the shifted 
and the unshifted particle (i.e., between a debris star and the cluster) 
is small, the relation between $\delta\varphi$ and the time $\Delta t$ since 
the release of the shifted particle can be expressed in a simple formula.
Provided that $\delta\varphi$ is small enough to ensure that the 
galactocentric distance $\alpha R$ along the orbit of the shifted particle 
and hence its angular velocity, which is $L / \alpha R^2$, can be 
considered as being approximately constant over $\delta\varphi$, it follows 
that 

\begin{eqnarray}
\delta\varphi \approx \frac{L}{\alpha R^2} \delta t = 
\frac{(\alpha-1)}{\alpha} \Delta t \frac{L}{R^2}\ .
\end{eqnarray}

Here, $\delta t$ means the time lag that corresponds to $\delta\varphi$,
for which equation (6) yields $\delta t = (\alpha - 1)\Delta t$.
Note that equation (7) is independent of the value of the circular 
velocity $v_c$ of the potential. The relation shown in equation (7) 
is very useful because it provides a key for estimating the mass loss rate 
of the cluster (see \S 6).

\subsection{Local orbit and tangential velocity} 
We now describe what observational constraints we have on the cluster's local 
orbit, i.e., its orbit near the present position of the cluster and the 
location of its tails.   
Adopting $d=23.2$~kpc (Harris 1996) for the heliocentric distance of 
Pal\,5 and $R_\odot = 8.0$~kpc for the distance of the Sun from the Galactic 
center we derive the position of Pal\,5 in the Galaxy as $(x,y,z) = 
(8.2,0.2,16.6)$~kpc. Here, $x,y,z$ denote right-handed galactocentric 
cartesian coordinates, with $y$ being parallel to the Galactic rotation of 
the local standard of rest (LSR) and $z$ pointing in the direction of the 
northern Galactic pole. In other words, the Sun has coordinates 
($-$8.0,0.0,0.0) in this system.
From the above position of Pal\,5 it follows that the inclination between the 
line of sight and the orbital plane of the cluster must be $\le 18^\circ$. 
On the other hand, our view of the orbital plane cannot be entirely edge-on 
because Figure 3 clearly shows the S-shape bending of the tidal debris near 
the cluster. This feature obviously reflects the opposite radial offsets 
between the two tails and the cluster. Considering the orientation of this 
S feature and the perspective of the observer, we infer that the orbit of 
the cluster (in projection on the plane of the sky) must be located east of 
the northern tail and west of the southern tail (referring to the equatorial 
coordinate system used in Fig.3). 

The simple model from \S 5.1 tells us that the tidal debris should 
be on similar orbits as the cluster if velocity differences can be neglected.
Taking into account the local symmetry of the tidal field, the limited range 
in azimuth angle $\varphi$ covered by the observations, and the relatively 
small angle between the orbital plane and the line of sight, 
one thus expects the offsets between the tails and the orbit of the cluster 
in projection on the tangential plane of the observer to be constant 
and of equal size on both sides of the cluster. An additional argument 
for this assumption is that the tails show a constant width, i.e., the 
projection does not reveal that they become wider as a function of angular 
distance from the cluster. 
If the mean (projected) separation between the tidal debris and the orbit of 
the cluster were increasing with angular distance from the cluster one should 
expect to see the tails to become wider, which is not the case.   
Therefore we continue the analysis under the assumption that the cluster's 
projected orbit runs parallel to the two tails.

First of all, this sets a tight constraint on the direction of the cluster's 
velocity vector in the tangential plane. The tails imply that the tangential 
motion of the cluster has a position angle of $231^\circ\pm 2^\circ$ with 
respect to the direction pointing to the northern equatorial pole, and 
$280^\circ\pm 2^\circ$ with respect to Galactic North (see Fig.~8). 
The orientation of this angle (i.e., $PA = 280^\circ$ and not 
$PA = 100^\circ$) follows when taking into account the direction to the 
Galactic center. Figure~9 shows the surface density map of the tails on  
a grid of galactic celestial coordinates $l\cos b,b$, where $l$ means 
galactic longitude and $b$ galactic latitude. 
Since the Galactic center ($l=0,b=0$) lies to the bottom of this plot, 
the tail that points to the right (also called the southern tail), must be 
the one at smaller galactocentric distance, which is thus leading, and the 
tail that points to the left (also called northern tail) be the more distant 
one, which trails behind. This means that the cluster is in prograde rotation 
about the Galaxy, in agreement with indications from different measurements 
of its absolute proper motion (Schweitzer, Cudworth \& Majewski 1993; 
Scholz et al.\ 1999; Cudworth 1998 unpublished, cited in Dinescu et 
al.\ 1999).

Next we consider whether the observed part of the stream is long enough to 
see a deviation from straight line motion. Figure~9 demonstrates that this 
curvature is indeed detectable. The long dashed line shows the projection of 
a straight line in space (with position angle $280^\circ$) plotted over the 
surface density contours of the debris.
The curvature of this line (due to projection onto the $(l\cos b,b)$ 
coordinate grid) is obviously too small to fit the tidal stream. Thus the 
curvature of the stream provides clear, direct evidence that the motion of 
the cluster is accelerated. 

To further constrain the orbit, the radial velocity of the cluster in the 
Galactic frame is needed and an assumption on the acceleration field near the 
cluster has to be made. The cluster's heliocentric radial velocity of $-58.7 
\pm 0.2$\,\kms\ (Paper~II), combined with a solar motion of 
$(U,V,W)_\odot =(10.0,5.3,7.2)$~\kms\ (Dehnen \& Binney 1998b, velocity 
components in our $x,y,z$ system), and $v_{\mathrm{LSR}} = 220$~\kms\ for the 
rotation velocity of the local standard of rest yields a Galactic rest frame 
radial velocity of $-44.3$~\kms (observer at rest at the present location of 
the Sun). 
The cluster's absolute proper motion is not yet measured with comparable 
accuracy. However, the existing measurements (see the above references)
can be used to derive rough limits for the tangential velocity.
According to Cudworth's measurement, which we consider to be the most 
reliable one, the absolute proper motion of Pal\,5 lies between 2.6 and 3.8 
mas/yr (3-$\sigma$ limits). 
Using the above values for the cluster's distance and the motion of the Sun 
and assuming the direction of the tangential velocity to be near $PA = 
280^\circ$, this yields a lower and upper limit for $v_t$ (i.e., tangential 
velocity as seen by observer at rest at present location of the Sun) 
of 60~\kms\ and 195~\kms, respectively.
To determine possible space velocity vectors for the cluster we thus
combined the radial velocity with tangential velocities in the 
range from 50~\kms\ to 200~\kms and the direction $PA =280^\circ$.

The simplest way to model the local acceleration field near the position 
of the cluster is a spherically symmetric field with constant acceleration. 
Assuming that the circular velocity in the Galactic halo is between 
150~\kms\ and 250~\kms, it follows that plausible values for the acceleration 
$a$ range from $(150\,\kms)^2/R_{cl}$ to $(250\,\kms)^2/R_{cl}$, with 
$R_{cl}$ being the cluster's galactocentric distance.
We thus adopted $a = (220\,\kms)^2/18.5\,\mathrm{kpc}$ and used this to 
integrate the orbit locally for a sequence of tangential velocities covering 
the interval from 40 to 180\,\kms\ in steps of 5 to 10\,\kms. Each orbit 
(more precisely its projection on the plane of the sky) was then compared 
with the tidal tails. 
It turned out that an orbit with a good fit to the geometry of the tails is 
obtained when using $v_t = 95\,\kms$. This orbit is shown by the solid line 
in Figure~9.
Changing $v_t$ by $\pm 15\,\kms$ leads to orbits with significantly different 
curvature (short-dashed and dot-dashed line in Fig.~9) which fit the tidal 
tails less well. These cases are considered as the limits of the range of 
acceptable orbits.
For other values of $a$ in the above range one can obtain orbits with 
essentially the same projected path when increasing or decreasing $v_t$
accordingly. This means that fitting a local orbit to the tidal tails 
sets up a relation between the acceleration $a$ and the cluster's 
tangential velocity $v_t$. The value of $a$ can however not be determined 
in this way from the current data unless the velocity $v_t$ is known 
independently. 

A more realistic model for the Galactic field near the cluster is given by 
the spherical logarithmic potential $\Phi = v_c^2 \ln R$, which yields a 
constant circular velocity $v_c$ rather than constant acceleration. We 
repeated the orbit integration using this model and again looked for the 
best accordance with the geometry of the tidal tails.
It turns out that for similar sets of parameter values the orbits obtained 
with this potential have practically the same projected paths as those 
obtained with the $a=const$ model. 
This implies that local variations in the size of the acceleration vector 
have little influence on the determination of the cluster's local orbit. 
Hence it does not matter which of the two models we use.  

The relation between the circular velocity $v_c$ of the logarithmic potential
(or the local acceleration $a=v+c^2/18.5$\,kpc) and the tangential velocity 
$v_t$ of the cluster, for which one obtains a local orbit with a projected  
path identical to the solid line in Figure~9, is shown in Figure~10. This 
relation is linear with a slope $dv_t/dv_c = 0.43$. A straightforward way 
to determine the parameter $v_c$ of the potential would be to obtain $v_t$ 
through a precise astrometric measurement of the cluster's absolute proper 
motion. An accuracy of 5 \kms\ in $v_c$ would require an accuracy of 2 \kms\ 
in $v_t$, which at $d=23.2$~kpc corresponds to a proper motion error of 18 
$\mu$as/y. This level of accuracy can be achieved in future astrometric space 
missions like SIM and GAIA . On the other hand, from the currently available 
proper motion measurements for Pal\,5 one can clearly not derive a meaningful 
constraint on $v_c$. The proper motion obtained by Cudworth yields 
$v_t = 135$ \kms and would thus suggest $v_c > 300$ \kms, which is 
uncomfortably high. The quoted error of this proper motion of 0.17 mas/y 
per component results in a typical uncertainty of 19 \kms\ in $v_t$, so that
$v_c$ cannot be determined to better than $\pm$47 \kms\ from this data.
A more comprehensive way of constraining the Galactic potential however 
consists in gathering kinematic data all along the tails and not only for the 
cluster. This will allow to use more complex models for the potential and to 
determine more than one parameter (e.g., the size and the direction of the 
local acceleration vector).     

In spherical potentials, the simplest motion would be that on a circular 
orbit. However, there is no such solution in the above sequence of orbits 
because for the given position, radial velocity, and direction of tangential 
motion a circular orbit would require a tangential velocity of 
$v_t =885.5\,\kms$, the total velocity on this orbit then being 
$v_c=886.6\,\kms$. 
Apart from the fact that velocities of this size are very far from realistic, 
it turns out that the projection of the resulting orbit would be similar
to that of motion on a straight line (see Fig.9) and hence not yield a good 
fit to the tidal tails. 
In order to allow a circular orbit with a velocity of 220\,\kms\ or less 
one needs to increase the position angle of the tangential velocity to 
$PA \ge 310^\circ$. However, the orbit would then deviate from the direction 
of the tails by at least $30^\circ$. A circular or nearly circular orbit is 
thus not an option.

We also checked for possible effects from a flattening of the Galactic 
potential. The potential in the Galactic halo is in fact likely to be 
flattened because of the influence of the disk, and may have additional 
flattening due to a flattened distribution of mass in the halo.
We thus constructed a model composed of (a) an exponential disk with scale 
length $h_r = 3\,\mathrm{kpc}$, scale height $h_z = 0.3\,\mathrm{kpc}$, and 
mass density $0.12\,M_\odot\mathrm{pc}^{-3}$ near the solar circle, and (b) 
a modified logarithmic halo potential 
$\Phi_h = v_c^2 \ln ((r^2+(z/q)^2)^{1/2})$ with $v_c = 180\,\kms$.
Here, $r=(x^2+y^2)^{-1/2}$ denotes the cylindrical radius. 
Together, the two components yield a flat rotation curve (in the 
Galactic plane) with a velocity of about 220~\kms.
In the region near the cluster the contribution of the disk to the 
total potential can be described with only a monopole and a quadrupole 
term. For the halo component two cases were considered, a spherical 
halo ($q=1.0$) and a flattened halo with $q=0.8$. 
Again, we find that these potentials lead to orbits whose local 
projections are practically indistiguishable from those obtained with 
the previous models. For $q=1.0$ the best-fit projected local orbit 
(in the sense of the solid line in Figure 9)
requires $v_t = 90\,\kms$. In the case of $q=0.8$ an equivalent orbit, 
i.e., with the same local projected path, can be obtained by increasing 
the tangential velocity to $v_t = 105\,\kms$.         
This demonstrates that over the currently known angular extent of the tidal 
tails the projected orbital path of the cluster is not sensitive to the 
flattening of the Galactic potential. 

In conclusion, the fit of the cluster's projected orbital path to the 
geometry of the tidal tails does not depend on a particular model for the 
Galactic field. 
As long as there are no additional kinematic constraints from radial 
velocities or proper motions along the tails the result of the fit is 
compatible with a variety of orbits for different fields, which locally 
project onto the same path on the sky. 
Those orbits of course differ from each other along the line-of-sight. 
However, within the region where we see the tidal tails the differences are 
not substantial. This allows us to estimate how much the distance 
varies along the cluster's orbit over the $10^\circ$ arc of the stream.
It turns out that the end of the leading tail at $l \cos b = -3\fdg2$\ 
lies nearest to the observer, in accordance with the negative radial 
velocity of the cluster.
The different orbits obtained with the above selection of Galactic models 
put this point at a galactocentric distance $R$ between 17.6 and 17.9 kpc, 
and at a heliocentric distance $d$ of 22.1 to 22.4~kpc. 
For the opposite end of the trailing tail at $l \cos b = +7\fdg0$\ 
these orbits predict galactocentric distances between 18.6 and 19.0~kpc, 
and heliocentric distances of 23.4 to 23.9~kpc.
The maximum of $d$ lies between 23.6 and 23.9~kpc. 
The distance between the cluster's orbit and the observer thus varies by 
up to 1.8~kpc over the length of both tails. This is a variation of +3\% 
and $-$5\% relative to the present distance of the cluster. Observationally, 
this corresponds to a magnitude difference of +0.06~mag and $-$0.1~mag, 
respectively, which is completely consistent with the results presented in 
\S 4.4. 

The maximum value of $R$ along the local orbit lies in the range from 18.7 
to 19.4~kpc (see upper panel of Fig.10). All solutions for the local orbit 
place its apocenter between 4\fdg1 and 6\fdg6 in $l \cos b$. 
This reveals that the cluster's present position must be close to the 
apocenter of its orbit.
The different orbit solutions suggest that the cluster passed this 
apogalactic point between 13 and 27 Myrs ago. The cluster's proximity to 
an apogalactic point implies, that the variation of the galactocentric 
distance $R$ along the local orbit is small. Indeed, in the region of 
the trailing tail the variation of R along the orbit is at most 0.6~kpc 
or 3\%, and in the region of the leading tail this variation is 0.9~kpc 
or 5\% (see the above minimum and maximum values).

\subsection{Is the tail aligned with the orbit?}
With a model of the cluster's orbit at hand, we can a posteriori test 
and validate our working hypothesis that the tidal tails lie parallel to 
the local orbital path of the cluster even if the stars do not have exactly 
the same velocity as the cluster when they decouple from the cluster. 
To this end we simulated a sample of test particles in a spherical 
logarithmic potential with $v_c=220$~\kms. The particles were released 
over the time interval from $-$2.0 Gyr to present at equal time steps 
of 20 Myr. We emphasize that this experiment is not meant to provide a 
realistic model of the mass loss history of the cluster, but just serves to 
reveal the geometry of the tails.
The particles were released from the cluster with a radial offset, i.e.,
either in the direction of the Galactic center or in the opposite direction, 
and with small velocity offsets. The size of the radial offset was chosen to 
be $3 \times r_L$, with $r_L$ the distance of the Lagrange point of local
force balance between the cluster and the Galactic potential, i.e., 

\begin{eqnarray}
r_L^3 = G M_{cl} \frac{R^2}{v_c^2} 
\end{eqnarray}

For the present position of the cluster equation (8) yields $r_L = 57$~pc,  
using $M_{cl} = 6 \times 10^3 M_{\odot}$ as the mass of the cluster.
The velocities of the particles were offset from the velocity of the cluster 
by a velocity vector with a size of 1.0~\kms\ (i.e., about $2\times$ the 
dynamical velocity dispersion of the cluster, see Paper II), 
pointing either in the radial direction (i.e., towards the Galactic center 
on the inner side and away from it on the outer side) or at $45^\circ$ from 
this direction.

Figure~11 shows the distribution of this sample of test particles in the 
plane of the sky at $t=0$. The plot covers the same field as Figure~9 and 
uses the same (galactic) coordinates. The solid line shows the path of the 
cluster's orbit, same as in Figure~9.
It can be seen that the stream of test particles has approximately the same 
width as the observed tails, and that it is well aligned to the cluster's
projected orbital path. In other words, the relatively small peculiar 
velocities that stars may have when escaping from Pal\,5 do not have a 
significant impact on the mean location of the tidal debris with respect to 
the cluster's orbit, at least not in projection onto the plane of the sky. 
Thus our assumption, that the orbit of the cluster must be fit such that it 
is parallel to the observed tidal tails proves to be entirely valid.

\subsection{Global orbit}

We saw that the apogalactic distance $R_{max}$ of Pal\,5 can be derived 
from the local orbit and hence does not strongly depend on specific 
assumptions on the Galactic field. 
However, the determination of other characteristic parameters of the 
cluster's orbit such as the perigalactic distance $R_{min}$ or the distance 
$R(z=0)$ at which the cluster crosses the Galactic disk requires extensive
extrapolation beyond the region of the tails and thus depends on a global 
Galactic mass model. In order to estimate these and other orbital 
parameters we made use of Model 2 from the series of Galactic models developed
by Dehnen \& Binney (1998a). This mass model consists of three exponential 
disks representing the stellar thin and thick disk and the interstellar 
material, and of a bulge and a halo component. The parameters of the model 
are chosen such that the model accomodates a variety of observational 
constraints, e.g., the Milky Way's rotation curve, the local vertical 
force, the local surface density of the disk etc. (for details see Dehnen 
\& Binney 1998a). 

The tangential velocity of Pal\,5 was set to $v_t = 90$\,\kms. Hereby the 
above model provides an orbit whose path locally coincides with the solid 
line in Figure~9 and hence meets the condition of a good fit to the tidal 
tails . 
The equations of motion were integrated over the time interval from $-$1~Gyr 
to 1~Gyr. Part of the resulting orbit is shown in Figure~12abc. 
We find that the orbit has perigalactic distances in the range from 6.7 to 
5.7~kpc.\footnote{
Note that for an orbit in a flattened potential perigalactic passages 
do in general not occur all at the same distance.}  
This reveals that the cluster penetrates deeply into the inner part of 
the Milky Way. 
The typical time scales of the orbit are $<T_R> = 291$\,Myr (mean period 
of radial oscillation) and $<T_{\psi}> = 443$\,Myr (mean period of 
rotation around the Galactic z-axis). 
The local constraints from the tidal tails allow us to vary $v_t$ by 
about $\pm 10$\,\kms.  
When doing so the perigalactic distances of the orbit change by about 
$\pm 0.8$\,kpc, and the eccentricity thus varies by $\pm 0.05$. The 
periods $<T_R>$ and $<T_\psi>$ change only slightly, i.e., by 
$\pm 3\%$ and $\pm 1\%$. 

From $-$1 Gyr to present the orbit makes five disk crossings. Three of them 
happen near or inside the solar circle, at distances of 6.7, 6.8, and 
8.3~kpc, while two are at much larger distances of 14 to 18~kpc. 
When changing $v_t$ by $\pm 10$\,\kms\ the galactocentric distances 
of the crossings of the inner disk vary by typically $\pm 0.7$\,kpc but 
occasionally up to $\pm 1.1$\,kpc. 
Interestingly, the predicted location of the next future disk crossing is 
at an even lower galactocentric distance of $5.9 \pm 0.8$\,kpc, which is  
very close to the next perigalacticon (see Figs.12a and 12b). 
This disk crossing is predicted to happen in $+110 \pm 2$\,Myr from present. 

Besides the models of Dehnen \& Binney there exist a variety of other 
Milky Way mass models from other authors, e.g., Pacynski (1990), 
Allen \& Santillan (1991), Johnston et al.\ (1995), Flynn et al.\ (1996).
To test in how far our conclusions on the orbit of Pal\,5 depend on the 
particular model we repeated the integration of the orbit using the same 
initial velocities and the Milky Way mass model of Allen \& Santillan. 
This is a three-component model with bulge, disk, and halo, where the disk 
potential is of the Miyamoto-Nagai form (Miyamoto \& Nagai 1975). 
The corresponding orbit is shown in Figure~12def. The general characteristics 
of the orbit are very similar to the one found with the Dehnen \& Binney
potential. In particular, we find small pericentric distances down to 
5.5~kpc. This confirms that the orbit of the cluster leads through the inner 
part of the Milky Way. 
The sequence of near and far disk crossings and apogalactic and perigalactic 
passages is the same as with the other model, but the associated time scales 
are somewhat shorter (e.g., $<T_R> = 275$~Myr, $<T_\psi> = 412$~Myr). 
Again, the orbit predicts an exceptionally small galactocentric distance, 
namely of 5.7\,kpc for the next crossing of the disk (in about 107~Myrs from 
present). When using a simple spherical logarithmic potential with $v_c$ 
in the range from 150 to 260\,\kms\ the resulting orbits yield even lower 
values for the galactocentric distance of this disk passage. It thus appears 
that an upper distance limit of $R \le 6$\,kpc for the next disk crossing 
is a safe prediction.

\section{Clues on the mass loss history}

\subsection{Mean mass loss rate}

Using the results from \S5 we can translate the amount of mass that is 
observed in the tails of Pal\,5 into a rough estimate of the mean mass loss 
rate. It was shown that the variation of $R$ along the local orbit is small, 
in particular in the region of the trailing tail. This means that the 
variation of the angular velocity $\dot{\varphi} = L / R^2$ along the local 
orbit is also small. 
Assuming $L=const$, the relative change in $\dot{\varphi}$ is twice the 
relative change in $R$ and thus  $\le 6\%$ for the trailing tail and 
$\le 10\%$ for the leading tail. Therefore, it is justified to estimate the 
time scale of the angular drift between the debris and the cluster in the way 
described by equation (7). 

The key parameter is the relative radial offset $\alpha$ from the orbit of 
the cluster. In reality, individual stars do not escape from the cluster 
under exactly the same conditions and hence do not settle on orbits with the 
same radial offset. Their orbits will not even be strictly confocal because 
they do not escape with exactly the same velocity. Hence stars at a certain 
azimuthal distance from the cluster will have taken different intervals of 
time to drift to this place. However we assume, that we can estimate the mean 
time scale $\Delta t$ of this drift by applying equation (7) to the mean 
value of $\alpha$. 

To determine the mean offset we measured for each star its rectangular 
separation from the solid line of Figure~9 in the plane of the sky. We then 
counted the weighted number of stars in $2'$ wide bins of this rectangular 
separation, using the same weighting scheme as described in \S3.2.  
Separate counts were made for the leading and the trailing tail. The resulting 
star count histograms are plotted in Figure~13. Each tail shows up as a 
symmetric peak on top of a constant background.
We determined the center and the width of each peak by fitting a Gaussian
plus a constant to the counts (see dashed lines in Fig.~13). For the trailing 
tail we thus measure a mean rectangular separation of 11\farcm8$\pm$0\farcm5 
from the orbit and a FWHM of 18\farcm4$\pm$1\farcm2. For the leading tail we 
find a mean separation of 10\farcm1$\pm$0\farcm8 and a FWHM of 
17\farcm2$\pm$1\farcm9. \footnote{When cutting the tails into two parts of 
equal length we get the same FWHM for each part within the errors of the fit. 
The width of the tails can thus be regarded as constant.}
From the mean angular separations as seen in projection we reconstructed the 
mean radial distance between the debris and the orbit of the cluster in the 
orbital plane. This was done in the following way: We increased and decreased 
the length of the galactocentric radius vector of the cluster by 200\,pc, 
determined the positions of the endpoints of these vectors on the sky as 
observed from the Sun, and then computed the rectangular separation of these 
points from the projected orbit in the same way as done for the stars.
This yields separations of 9\farcm0 and 9\farcm1, respectively, i.e., 0.76 
and 0.90 times the observed separations. This implies that the observed 
separations correspond to mean radial distances between the tails and the 
orbit of 263\,pc for the trailing tail and 222\,pc for the leading tail.
  
We first focus on the trailing tail, which is better covered by the 
observations and which is most suited for applying equation (7). 
Using $v_t=95$\,\kms\ the angular momentum of the cluster's orbit is 
$L= 1814$\,\kms\,kpc . From $R_{max}= 19.0$\,kpc we derive the angular 
velocity at the apogalactic point as $\dot\varphi = 5.14$\,\kms/kpc or, 
in other units, $\dot\varphi = 0\fdg295$/Myr. 
Comparing the mean radial offset of the tail of 263\,pc and the apogalactic 
distance of 19.0\,kpc we have $\alpha = 1.014$.  
The material in the trailing tail is basically spread over an arc of 
$6^\circ$ on the sky. Along this arc, the orbit of the cluster subtends an 
azimuth angle of 7\fdg7 in the orbital plane. 

Putting these numbers into equation (7) we obtain $\Delta t = 1.94$\,Gyr. 
This is the typical time it has taken debris stars to drift from the cluster 
center to the ``tip'' of the tail. 
The trailing tail contains about 0.8 times as many stars as the cluster 
(see \S4.1). If we assume that the tail has the same mass function as the 
cluster, the mass in the tail should be $0.8 \times M_{cl}$, where $M_{cl}$
denotes the present mass of the cluster. With regard to the mass function, 
this is a lower limit, because the tail is likely to contain a larger 
fraction of low-mass stars than the cluster. The cluster is known 
to be underabundant in low-mass stars and may have lost them through mass
segregation followed by tidal stripping from the outer part of the cluster. 
Since low-mass stars are not represented in our sample such a difference in 
the mass function would result in a somewhat higher total mass of the tail. 
Using the above values, and taking into account that equal amounts of mass 
are lost on both sides of the cluster, we finally obtain an estimate of the 
mean mass loss rate of $-\dot{M}/M_{cl} = 0.82$/Gyr. Multiplying by the 
present mass of the cluster, which has recently been estimated to be $-M_{cl} 
= 6 \times 10^3 M_{\odot}$ (Paper II), we get $\dot{M} = 4.9 M_{\odot}$/Myr.  

To check this result we do an analogous calculation for the leading tail
using a mean distance of $R=18.2$\,kpc (see \S5.2). The mean angular 
velocity then is $\dot\varphi = 0.321^\circ$/Myr, and the mean radial 
offset of the tail from the orbit of the cluster yields $\alpha =0.988$.
Furthermore we have $\delta\varphi = 4\fdg8$ as the azimuth angle of the 
orbit along the leading tail (seen from the Galactic center), and 
$M_{tail} = 0.4 M_{cl}$ as an estimate of its mass (see \S 4.1). 
Equation (7) thus yields $\Delta t = 1.26$\,Gyr, and this leads to a mean 
mass loss rate of $-\dot{M}/M_{cl} = 0.63$/Gyr or 
$-\dot{M} = 3.8 M_{\odot}$/Myr. 
This rate is somewhat lower than the one obtained from the trailing tail
because the mean drift rate along the leading tail is only slightly higher
and cannot compensate the lower surface density in the leading tail. 

The accuracy of these estimates is limited by a number of potential sources 
of errors, the most important of which are: 
(1) The uncertainty in the determination of the offset between the tails and 
the orbit. (2) The uncertainty of the angular velocity resulting from errors 
in the distance and the tangential velocity of the cluster. (3) Deviations of 
the motion of individual stars from the drift rate at the mean radial offset 
due to a spread in their initial positions and velocities.
While the fitting of the histograms of Figure~13 yields a formal uncertainties of 0\farcm5 and 0\farcm8 for the mean angular separation, we expect that the 
true error of this quantity is more like $1'$ to 1\farcm5, in particular 
because the exact location of the cluster's orbit is not known. 
The relative error in the mean angular separation and hence in the mean 
radial offset is thus assumed to be between 10\% and 15\%, producing a 
relative error in the factor $(\alpha-1)/\alpha$ of approximately the same 
size. The heliocentric distance and the estimate of the tangential velocity 
of the cluster are both thought to have a relative uncertainty of 10\%.
This translates into relative errors of 12\% and 10\% for the corresponding 
galactocentric quantities and thus results in a combined error for the
angular velocity $\dot\varphi$ of 16\%. Hereby, the relative error in the 
fractional mass loss rate would be about 22\%. How much the above estimate 
of the mass loss rate is biassed by the spread in the drift motions of 
individual stars needs to be investigated with forthcoming detailed N-body 
simulations of the tidal disruption Pal\,5. 
However, we expect that this may also contribute a relative uncertainty of 
about 20\%. The total error is thus believed to be of the order of 30\%. 
In conclusion, our result for the mean mass loss rate of the cluster 
with respect to its present mass is $-\dot{M}/M_{cl} = 0.7 \pm 0.2$/Gyr 
or $-\dot{M}=4.3\pm1.3 M_\odot$/Myr (average from both tails).

It is interesting to compare this mass loss rate with predictions from 
approximate formulae devised by Johnston et al.\ 2002 (hereafter JCG02) for 
calculating mass loss rates of satellites from radial surface density 
profiles. These formulae (see equations (6) and (7) of JCG02) relate the 
mass loss rate to the break radius of the surface density profile, the 
surface density outside this radius, and the time period for orbital or 
circular motion, assuming that the radial profile decreases as either 
$r^{-1}$ or $r^{-2}$. Since the observed profiles for Pal5 are characterized 
by $-1.5 \le \gamma \le -1.2$ they lie in between those two cases. 
We calculated the predicted mass loss rates with two different radii, 
$r=20'$ and $r=60'$. Using $r_{break}=16'$, average surface densities as 
provided in Figure~5, and the time scales $T_{orb} = 356$~Myr, $T_{circ} = 
528$~Myr, equation (6) of JCG02 predicts mass loss rates of $6.2 M_\odot$/Myr 
($with r=20'$) and $3.7 M_\odot$/Myr (with $r=60'$) while equation (7) of 
JCG02 predicts mass loss rates of $2.1 M_\odot$/Myr and $3.8 M_\odot$/Myr, 
respectively. The predictions are hence within 50\% of our above detailed 
estimate of the mean mass loss rate of the cluster.

\subsection{Total mass loss}
According to numerical simulations the mean rate of tidal mass loss of a 
globular cluster orbiting in a stationary Galactic potential remains 
approximately constant over most of its lifetime, i.e., the cluster's mass 
decreases approximately linear with time (see, e.g., Gnedin et al.\  1999, 
Johnston et al.\ 1999a, Baumgardt \& Makino 2002). 
From the observations it is not entirely clear whether this really holds 
for Pal\,5 because the outward decrease of the linear density of the tails 
could be interpreted as an indication for a secular increase in the mean 
mass loss with time.
However, if we assume that the mean mass loss rate was constant, we can use 
the measured mean rate of recent mass loss to get an idea of the cluster's 
mass at earlier epochs. An important condition is that the Galactic 
potential must have been essentially the same over the time interval spanned 
by the extrapolation. Extrapolating over the cluster's entire age of 12 Gyr 
thus poses a problem because disk shocks seem to be an important driver of 
the mass loss of Pal\,5, and the age distribution of disk stars suggests 
that the Galactic disk was not in place at very early epochs. However, if the 
disk was less massive or non-existent early on, the initial rate of the 
cluster's tidal mass loss was presumably lower than the present rate.
Therefore, an extrapolation with the present mass loss rate may overestimate
the cluster's initial mass but can put an upper limit on it.

Our measurements suggest a mean mass loss rate of $-\dot{M}/M_{cl} = 0.7$/Gyr.
Multiplying this with 12 Gyr and adding the cluster's present mass, we obtain 
an upper limit for the cluster's initial mass of $M = 9.4 M_{cl}$. 
Pal\,5 may thus have started with roughly ten times as much mass as it has 
today.
If one restricts the extrapolation to a time interval of about 8 Gyr, for 
which the existence of a massive Galactic disk is likely, one obtains a 
cluster mass of $6.6 M_{cl}$. Since the cluster is certainly 
older than 8 Gyr, this presents a lower limit to its initial mass, provided 
that the cluster has been on its present orbit during the entire period. 
Of course, these limits vary as a function of the error of the mass loss rate
and provide not more than a rough guide line.

\section{Discussion and Summary}

\subsection{Compelling evidence for tidal disruption}
Our analysis of an enlarged set of SDSS data reveals that the cluster 
Pal\,5 is connected to a long stream of tidal debris that contains at least 
1.2 times as much stellar mass as the cluster itself contains in its present 
state. This confirms and extends the results of Paper~I, which presented 
first evidence for tidal tails from a study of a smaller field. 
The most basic aspect of these results is that they provide the first 
stringent observational proof that globular clusters in the Milky Way's 
halo may be subject to significant tidal mass loss, by which they eventually 
dissolve.
The detection of fully-fledged tidal tails, which is unique so far, makes
Pal\,5 the prototype of such tidally-disrupting globular clusters. 
We showed that the stream of debris is thin and maintains its small width 
over a length of several kiloparsecs, suggesting that it is a kinematically 
cold system. This is very much consistent the low velocity dispersion inside 
the cluster, which is only about 0.5\,\kms\ (Paper II).  
The numerical experiment described in \S 5.3 and Figure~11 lends further 
support to this view.  It is obvious that the relatively small transverse 
spread of the debris on the sky has strongly favored the detection of the 
stream. 
In the case of a massive cluster, having a much higher internal velocity 
dispersion, 
the debris would probably spread out in a much wider stream and thus 
be more difficult to detect.  
Another favorable circumstance is the fact that Pal\,5 is 
presently located close to its apogalacticon. This means that the 
distribution of the debris has the smallest possible angular dispersion 
along the orbit and therefore shows a relatively high density.

\subsection{The tails and the orbit of the cluster}

It was demonstrated that the arc over which the tidal debris has been 
traced is now sufficiently long to recognize the intrinsic curvature of 
the stream. 
Since there is very good reason to assume that the tails are closely aligned 
with the orbit of the cluster the curvature of the stream reveals the local 
curvature of the cluster's orbit. This is a remarkable point because 
curvature means acceleration. Direct measurements of the Galactic 
gravitational acceleration of an individual halo star or star cluster in 
the sense of observing a non-linear change in position or a change in 
velocity over time are for technical reasons totally out of reach.  
Therefore the curvature, and also the bipolar and S-shaped structure of 
Pal\,5's stream of debris provides one of the first occasions where the 
acceleration of a halo object by the gravitational field of the Galaxy is 
directly visible. 
A considerable drawback however is that the tidal stream is known only in 
projection on the plane of the sky.  
This projected view provides clear evidence that the acceleration of the 
cluster is non-zero, but does not yet allow to derive a useful constraint 
on its specific value. To achieve this the observations would need to cover a 
major part of the cluster's orbit or one would require precise information 
on the tangential velocity of the cluster or on the variation of the distance 
or the kinematics along the tails. As an example, we determined the 
best-fitting orbits in a spherical logarithmic potential and showed that 
measuring the absolute proper motion of the cluster on the accuracy level of 
10 $\mu$as/y would allow to estimate the local circular velocity to about 2\%. This is similar to the conclusion drawn by Johnston et al.\ (1999b) from a 
study of simulated tidal streams.

Using different Galactic models and selecting orbits that optimally fit 
the tidal tails, we found that the cluster must be near its local maximum 
distance from the Galactic center and that therefore, the galactocentric 
as well as the heliocentric distance along its orbit varies by no more than 
a few percent over the length of the tails. 
This is in full agreement with the photometry, because color-magnitude 
diagrams show that there is no systematic difference in apparent brightness 
between the tidal debris and the cluster of more than 0.1 mag. 
On the other hand, it turned out that the pericenters of the cluster's orbit 
must lie in the inner Galaxy, namely at $R \le 7$\,kpc. 
This means that its orbit is rather eccentric and that the cluster must have 
repeatedly crossed the Galactic disk near or inside the solar circle. 
Since such disk crossings at small galactocentric radii lead to strong tidal 
shocks this provides a convincing explanation for Pal\,5's heavy mass loss. 

Of particular interest is the fact that the next disk crossing, which 
will happen in about 100~Myr, is predicted to take place at a galactocentric 
distance $\le 6$\,kpc, causing a very strong tidal shock. 
Given the small amount of mass that is left in the cluster and its low 
spatial concentration, one must suspect that this event will trigger the 
total disruption of the cluster. 
Dinescu et al.\ (1999), using formulae for the destruction rates due to 
disk- and bulge shocks as developed by Gnedin \& Ostriker (1997) and 
orbital parameters determined from measured proper motions, derived 
a theoretical estimate of the destruction time of 0.1 Gyr for Pal\,5. 
This is identical to our estimate for the time until the next disk crossing. 
On the other hand, the original paper by Gnedin \& Ostriker (1997) gave 
estimates of the destruction time for Pal\,5 of either 1.1 Gyr or 46 Gyr, 
depending on the model for the Galactic potential. 
These time scales are too long in view of what is known about Pal\,5 now, 
and must have resulted from a very different model for the cluster's 
orbit.\footnote{Note that these authors used random values for the 
cluster's velocity components perpendicular to the line of sight.} 
This demonstrates the importance of reliable individual orbital data for 
clusters like Pal\,5 since such data can place much stronger constraints on 
the dynamical evolution of these systems than purely theoretical analyses.

\subsection{The surface density in the tails}

The surface density distribution of the stars in the tails shows 
some important peculiarities: \\
(1) The distribution is clumpy, i.e., there are a number of density maxima 
and minima in each tail. While the density variations in the leading tail 
are moderate and not necessarily significant (i.e., they could still be the 
result of statistical fluctuations) the trailing tail shows at least two 
strong density maxima and one gap, where the local density deviates 
significantly from the mean density level. 
Although disk shocks, which are believed to be the primary cause of Pal\,5's
mass loss, are likely to modulate the instantaneous mass loss rate, it is 
implausible that the observed clumps are due to this modulation because the 
density variations would then need to occur symmetrically in both tails. 
Also, such variations should preferentially be visible near the cluster while 
at larger angular distances they should be washed out by the differential 
drift between stars of different orbital energy. 

The region of maximal stellar density in the trailing tail lies at arc 
lengths between 2\fdg3 and 3\fdg7 from the center of Pal\,5. According to 
the results from \S6.1 the mean rate of apparent drift of the tidal debris
is 0\fdg31/100~Myr. Thus the stars that form this broad density clump are 
expected to have escaped from the cluster in the interval between 740 and 
1190~Myr ago. In \S5.4 we found that during the last 1 Gyr the cluster 
crossed the Galactic disk five times, and three times thereof at small 
distances from the Galactic center. The latter occurred 140~Myr, 480~Myr, 
and 740~Myr before present. If these inner disk crossings produce 
overdensities that can be observed as distinct clumps in projection on the 
sky, then one would expect to see one such clump close to the cluster 
($\lambda \approx 0\fdg3$) and two clumps - or perhaps one broad clump 
from the merging of the two - at arc lengths between 1\fdg5 and 2\fdg3.
This does not correspond to the observations. In order to achieve some kind 
of agreement one would need to assume that the drift of the debris is in fact 
50\% faster than derived from the radial offset between the tails and the 
orbit of the cluster.

It may seem intriguing that the major density enhancement in the stream is 
near the apogalactic point of the cluster's orbit. Nevertheless, the fact 
that a stellar stream gets compressed near the apocenter 
(because the angular velocity is minimal at this point) cannot explain this 
local enhancement because the angular scale of the observed feature 
(less than 2\fdg5 when viewed from the galactic center) is much too small. 
One could speculate that the density variations in the tails might come from 
scattering by small-scale perturbations of the Galactic potential as produced 
by spiral arms, molecular clouds, dark matter clumps etc. However it has not 
yet been demonstrated that such perturbations can indeed generate the 
observed features. We note that such perturbations would certainly need to be 
much weaker than those in a dark halo of massive ($\approx 10^5 M_\odot$) 
black holes because encounters with such massive compact objects would have 
destroyed the cluster - and most likely also its tails - on a very short time 
scale (Moore 1993).

One might also think that the broad density enhancement in the trailing tail 
could be related to the cluster M\,5 since it is uncomfortably close to it on 
the sky. But a dedicated search for debris from M\,5 (using a proper 
color-magnitude filter for this cluster) yields absolutely no evidence 
for an extended distribution of M\,5 stars that could overlap with the tail 
of Pal\,5.    

(2) Apart from local variations, the radial profile of the stellar surface 
density at $r \le 20'$ declines like a power law with an exponent of 
$-$1.2 to $-$1.5, which means that the linear density along the tails 
decreases slowly with increasing distance from the cluster. 
How does this compare to the results of N-body simulations of stellar 
systems in Milky Way like potentials?
Combes et al.\ (1999) show density profiles for two of their 
simulated clusters and find that at radii $r > r_t$
the volume density of debris stars decreases like $r^{-4}$, hence the 
surface density decreases like $r^{-3}$. This is clearly a much steeper 
decline than the one we observe.  
In contrast to this Johnston et al.\ (1999a) show surface density 
profiles of simulated globular cluster-like systems, in which the unbound 
outer part has a much shallower decline that is almost like $1/r$. 
As these authors point out, such a $1/r$ decline can easily be explained 
if one considers the very simplified case of a cluster on a circular orbit, 
which looses mass at a constant rate and with constant energy offsets. 
Our observations are not extremely far from the simple $1/r$ case but do not 
match it exactly.   
More recently Johnston et al.\ (2002), presented a series of simulations, 
in which the obtained radial surface density profiles that show a wide 
variety of logarithmic slopes in the outer part. They report values for 
the power-law exponent $\gamma$ between $-$1 and $-$4, and show that the 
result depends to some extent on the parameters of the orbit and the 
orbital phase.   
Even for an almost circular orbit they find $\gamma$ to scatter between 
$-$1 and $-$3. For very eccentric orbits their simulations seem to predict 
that $\gamma$ is mostly below $-3$, in particular at the apocenter. However, 
the system parameters (mass,radius) and the orbits that were used in these 
recent simulations are actually more representative of dwarf satellites than
of clusters like Pal\,5. 
The conclusion from this comparison is that different simulations 
predict a wide range of possible power law exponents for the debris and 
that our observed values of $\gamma$ lie in the upper part of this range.

The fact that in both tails the observed surface density profile is somewhat 
steeper than the simple $1/r$ may indicate that either the angular velocity 
along the orbit increases with increasing angular distance from the cluster
or the (orbit-averaged) mass loss rate has undergone a secular change. 
The variation of the orbital angular velocity must certainly enter the game
on larger scales but cannot have a significant impact on the present 
results because the arc length of the tails is too short. 
We are thus left with the possibility that the mean mass loss rate may have 
steadily increased or that the mass loss process may have suddenly set in 
not much longer than 2 Gyrs ago. 
In the latter case an outward decrease of the linear density of the tails 
could result from the fact that the tip of each tail would only contain stars 
with the highest energy offset from the cluster, which should occur in low 
numbers, while at smaller angular distance from the cluster one would find 
both, stars with smaller energy offset, which should be more numerous, and 
also stars with higher energy offset, which were released from the cluster 
more recently.     
 
Indeed, the traces of the trailing tail disappear before the tail would reach 
the edge of the observed field, and it is unknown whether or not the stream 
continues and reappears with significant density farther out along the orbit. 
If so, there would be a gap or a section of very low density with an arc 
length $\le 1^\circ$, that would need to be explained. 
If not, this would imply that the mass loss history of Pal\,5 underwent an 
abrupt change about 2 Gyr ago and that the mass loss rate in the earlier 
phase was much lower than in the recent phase or even zero. This would 
certainly require a fundamental change of the cluster's orbit. 

One possibility would then be that Pal\,5 came from outside the Galaxy and 
was accreted by it in a merger event with a smaller galaxy. This merger event 
would need to have happened fairly recently, and the cluster would have been 
accreted as a system with a mass of only $1.5\times 10^4 M_\odot$. 
As we will discuss in \S 7.7, an association of Pal\,5 with the Sagittarius
(Sgr) dSph galaxy is very unlikely. 
From statistical studies there is evidence for further halo substructure, 
i.e., possible great circle streams of outer halo satellites (e.g., 
Lynden-Bell \& Lynden-Bell 1995, Palma et al.\ 2002), which might be a hint 
on former merger events.  
However, based on sky position, distance, and radial velocity, Pal\,5 has 
not qualified as a possible member of any hypothetical stream in these 
studies.   
Since not all of the satellites may in fact belong to such a stream, an
association with an individual satellite needs not necessarily be obvious 
through great circle alignments. In this sense the question of whether
or not Pal\,5 is likely to be associated with one of the dSph satellites 
remains open because there is currently no well-constrained orbital 
information on most of these objects.

As another possibility we mention that the orbit of Pal\,5 could have 
changed drastically through a close encounter with one of the Milky Way 
satellites. Since the present orbit of Pal\,5 is unlikely to reach
beyond galactocentric distances of 20~kpc, the only known candidate for such 
an encounter would be the Sgr dwarf spheroidal. 
However such scenarios are highly speculative. It is thus very important
to learn more about the spatial extent of the stream. 

(3) Overall, the stellar surface density in the leading tail is on a lower 
level than in the trailing tail. This is surprising because the symmetry of 
the tidal force field suggests that the distribution of tidal debris should
- at least in the vicinity of the cluster - be symmetric with respect to the 
cluster center. Both, orbit calculations and the analysis of the photometric 
data have shown that variations in the line-of-sight distances cannot be 
responsible for this effect. 
One may thus wonder if the star count results could be influenced by variable 
interstellar extinction. In Figure~14 we show the surface density contours of 
Pal\,5 and its tidal stream overplotted on a grey-scale map of interstellar 
extinction derived from the reddening data of Schlegel, Finkbeiner \& Davis 
(1998). Although integrated Galactic foreground extinction corrections do not 
necessarily hold for stars within the Milky Way because they are integrated 
along the entire line of sight, we believe that this only leads to small 
errors for a cluster as distant and as far above the plane as Pal\,5. 
Another possible effect is the dependence of extinction on stellar 
temperature (and thus color) as discussed in Grebel \& Roberts (1995). 
Considering our filter choice and color range, this should amount to at most 
0.02 to 0.03 mag uncertainty, which is negligible for our purposes. 

Figure~14 reveals that there is indeed significant variation in interstellar 
extinction over the field, as was already mentioned in \S2. One notices 
that the leading tail is located close to the region of enhanced extinction 
but fortunately does not run across this region. This shows that there is 
little reason to assume a significant impact of extinction on the measurement 
of the stellar surface density in the tails. 
If enhanced extinction played a role, it would result in a loss of faint 
stars close to the detection limit. While the definition of our sample 
already involves a magnitude cut-off at $i^*=21.8$ well above the limit, we 
tentatively increased the cut-off by an additional 0.5 mag, 
thus restricting the sample to $i \le 21.3$. It turned out that the resulting 
surface density distribution is similar to the one for the larger sample and 
maintains the same overall imbalance between the leading and the trailing 
tail. It is therefore clear that the overall difference between the number 
counts in the two tails is not the result of variable extinction.

\subsection{The luminosity functions}
We showed that the luminosity function of the stars in the tidal tails 
is in very good agreement with the stellar luminosity function in the 
cluster. This result is not self-evident, although it nicely fits 
the idea that the tails consist of stars from the same stellar population 
as the cluster. As briefly noted in previous sections, a deep study of the 
core of Pal\,5 using the HST (Grillmair \& Smith 2001) has revealed that the 
luminosity function of Pal\,5 is relatively flat. 
This means that there is a strong deficiency in low-mass stars, at least  
in the core of the cluster. If this deficiency is a consequence of tidal 
mass loss (in combination with mass segregation), as Grillmair \& Smith 
suggested, one would expect to find a corresponding overabundance of 
low-mass stars in the cluster's debris.  
The luminosity functions should then actually be different. However, the 
flattening of the luminosity function shown by Grillmair \& Smith becomes 
effective at absolute magnitudes $M_I > 5$\,mag which is below the limit 
of our analysis of SDSS data. Therefore, even though we find that the 
luminosity functions of the tails and the cluster agree down to the limit 
of this study, it seems likely that they will diverge when probing
the stellar content of the tails at fainter magnitudes. If so, the 
higher fraction of low-mass stars in the tidal debris will contribute 
additional mass to the tails. This means that our current estimate of 
the mass ratio between the tails and the cluster (\S4.1) provides a 
lower limit while the true mass ratio may be somewhat higher.

\subsection{The mass loss rate} 
We showed that the transverse offset between the tails and the cluster 
can be used to estimate the time needed by debris stars to drift away 
from the cluster by a certain angle. This leads to the conclusion that 
the tails in their currently known extent represent the cluster's mass 
loss from essentially the last 2 Gyrs. The result of our numerical 
experiment on the drift of debris along the cluster's orbit, which is shown 
in Figure~11, confirms this time scale. It follows that the mean mass loss 
rate of Pal\,5 in this period was about 0.7($\pm$ 0.2) times the present 
mass of the cluster per Gyr.     
However, these estimates involve a number of approximations and simplifying 
assumptions. In particular, it remains to be investigated whether the fact 
that stars are actually distributed over a range of velocities and positions 
when escaping from the cluster leads to a substantial bias in the above 
estimates.
For more definitive results accurate modelling of the details of the 
mass loss process are needed. Thus N-body simulations of the system under 
realistic conditions need to be performed.

A simple extrapolation with the present mean mass loss rate shows
that the cluster may initially have had a total mass of between 6 and 
10 times its present mass. Obviously, such an estimate depends on whether 
the cluster maintained the same mean (i.e., orbit-averaged) mass loss 
rate over most of its lifetime. For time intervals in which (1) the Galactic 
potential was the same as today and in which (2) the cluster followed the 
same orbit as it is today, one can indeed expect this to hold true. 
However, the observed decline of the surface density of the tails with 
increasing distance from the cluster warns us that the mean mass loss rate 
might in earlier times have been lower than it presently is. 
For clarifying this issue it is important to find out whether 
or not the tails extend to larger distances from the cluster. 

\subsection{Tangential velocity versus proper motion}
The condition that the local orbit of the cluster needs to fit the location 
and curvature of the tidal tails allowed us to determine the vector of the 
tangential velocity of the cluster in a completely new manner. It is 
interesting to see whether the tangential velocity obtained in this way is 
consistent with the measured absolute proper motion of Pal\,5. 
Dinescu et al.\ 1999 report the proper motion of Pal\,5 measured by 
Cudworth (1998, unpublished) as $\mu_\alpha \cos\delta = -2.55\pm0.17$\,mas/y 
and $\mu_\delta = -1.93\pm0.17$\,mas/y. 
The quoted proper motion error corresponds to 19~\kms\ per component at the 
distance of the cluster. If one transforms the above proper motion into the 
galactic rest frame using $d=23.2$~kpc and the velocity components of the 
Sun specified in \S 5.2 one obtains a tangential velocity of 137~\kms\ with 
position angle $PA = 300^\circ$.
Thus the direction of the cluster's tangential motion derived from the 
measured proper motion differs from the direction given by the 
tidal tails by $20^\circ$. The absolute velocities show a difference of 
about 40~\kms. Alternatively, one can do the inverse transformation, assuming 
a tangential velocity of $v_t = 90$\,\kms\ or $v_t = 95$\,\kms\ with position 
angle $PA = 280^\circ$ as implied by the fit of the local orbit. 
This yields predicted proper motions of 
$\mu_\alpha \cos\delta = -2.01$\,mas/y, $\mu_\delta = -2.03$\,mas/y, or 
$\mu_\alpha \cos\delta = -2.05$\,mas/y, $\mu_\delta = -2.06$\,mas/y, 
respectively. 
Comparison with the measured values shows that the declination component 
agrees well with our predictions within the quoted error while the right 
ascension component deviates from the prediction by about $3\sigma$ or 
0.5~mas/y. Although this looks like a significant difference between the 
two completely independent determinations of the cluster's tangential motion, 
a proper motion difference of 0.5~mas/y is actually not unreasonably large 
and could be explained by an underestimation of the measuring error of 
the proper motions.  \\

\subsection{A former member of the Sgr dwarf?}

Based on position, radial velocity and rough proper motion data, 
some authors have argued for a possible association of Pal\,5 with the 
Sgr dwarf galaxy (Lin 1996; Palma et al.\ 2002, Bellazzini et al.\ 
2003). Palma et al.\ classified it as a possible but unlikely member of Sgr, 
because the pole families were similar but orbital energy and angular 
momentum were found to be different. 
Bellazzini et al.\ argued that Pal\,5 lies relatively close to the orbit 
of Sgr as given by the model of Ibata \& Lewis (1998), both in position 
$(x,y,z)$ and in the plane of $v_r$ vs.\ $R$. 
The impression of a good agreement in radial velocity $v_r$ is however fake
because one gets this only by associating the cluster with a wrong position 
along the orbit of Sgr. 
In reality, the radial velocity of the orbit of Sgr near the position of 
Pal\,5 differs from the radial velocity of Pal\,5 by about 100\,\kms 
(adopting the model of Ibata \& Lewis). 
Thus there is some accordance in position but not in kinematics.

The tidal tails and the cluster's local orbit that we derive from it allow 
a much more robust comparison and strengthen the evidence against a former 
membership to Sgr. While the orbit of Sgr is almost polar, that of Pal\,5 
is clearly not so. Hence Pal\,5 does certainly not orbit in the plane of 
the Sgr stream. The local orbit of the cluster crosses the orbit 
of Sgr at a large angle (compare our Fig.~12 with Fig.~1 of Bellazzini et 
al., but note that the orientation of the y-axis is inverted). 
Thus not only the radial motion, but also the tangential motion of the 
cluster is clearly discordant with the orbit of Sgr.  
The space velocity vectors of Pal\,5 and of the orbit of Sgr enclose an 
angle of $108^\circ$. 
Another argument is the strong difference in apogalactic distance.  
We showed that Pal\,5 is almost at its apogalacticon and does certainly 
not reach galactocentric distances of more than 20~kpc. The orbit of the 
Sgr dwarf however is thought to have apogalactica between 50 and 60~kpc 
(see, e.g., Dinescu et al.\ 2000, Bellazzini et al.\ 2002). This reflects
a large difference in specific orbital energy. Thus, former membership 
in Sgr would require that the cluster lost much of its orbital energy after 
the departure from its host, which would be difficult to explain.   
This shows that apart from an approximate positional correlation with 
the orbit of Sgr, which can be a coincidence, there is little reason to 
assume a connection between Pal\,5 and the Sgr dwarf. On the contrary, 
our results on the cluster's orbit make it rather unlikely that Pal\,5 
originates from this dwarf. This also holds if one considers other 
models for the orbit of the Sgr dwarf, e.g., by Helmi \& White (2001), 
Johnston et al.\ (1999c), or Gomez-Flechoso, Fux \& Martinet (1999), 
since all of them are polar. However, this does not exclude the possibility 
of a close encounter of the two systems by which the orbit of Pal\,5 might 
have been deflected (see \S 7.3).

\section{Outlook}

The tidal stream of Pal\,5 opens a new and promising way to constrain the 
gravitational potential in the Galactic halo. Detailed information on the 
orbits of individual halo objects like Pal\,5 or the Sgr dSph from their 
tidal debris can in principle produce much more powerful constraints on the    Galactic potential than classic statistical approaches.
We showed that with the current positional data for the tails of Pal\,5, 
which cover a limited range of orbital phase angles, the orbit of the cluster 
is not yet uniquely determined by the observations alone,  and 
conclusions on the Galactic potential can therefore not yet be drawn. 
However, this will change drastically when kinematic data for the tails are 
added to the analysis (see, e.g., Murali \& Dubinski 1999). 
Precise proper motions could be very useful, but will remain unavailable
until future astrometric space missions like SIM or GAIA are flown because
very high accuracy is required. On the other hand, precise radial velocities 
for giants stars associated with Pal\,5 are within reach of today's 8-10m 
class telescopes. Another approach is to measure main-sequence turn-off 
stars, which are more numerous, but much fainter and hence can only be 
observed at lower spectral resolution.
An observing program to obtain radial velocities of candidate giants along 
the tails of Pal\,5 has been started on the VLT. 
We expect that the results of this program will allow us to break the 
degeneracy in the determination of the cluster's local orbit and allow
a direct measurement of the gravitational acceleration in the Milky Way 
halo at a galactocentric radius of 18 to 19~kpc. 

A major open question is how far the tidal stream continues and  
what the full time span of the mass loss history of Pal\,5 thus is. 
This can be clarified with targeted searches for further tidal debris 
from Pal\,5 along the arc outlined by our model of the cluster's orbit. 
If the stream can be traced farther out, one should at some point also 
discover a substantial variation in the heliocentric distance of the 
stars. Hence such detections would not only provide information on the
mass loss history, but also provide further important constraints on the 
orbit.

The fact that we see Pal\,5 while it is only about 100 Myrs away from its 
complete disruption, makes it very likely that there have been more clusters 
of similar type, which dissolved during the last few Gyrs. This provides 
observational support for the common conjecture that the Milky Way's globular 
cluster system was originally much richer in low-mass cluster than it is 
today (see Fall \& Zhang 2002 and references therein). 
Since the tidal stream of Pal\,5 is at least about 2 Gyrs old, tidal streams 
from other low-mass clusters that dissolved recently may also still exist and 
be observable. Without the presence of a parent object such streams are of 
course more difficult to find.
On the other hand, the detection of anonymous streams that are left-overs 
from globular clusters would provide important information on the evolution
of the globular cluster system of the Milky Way and also provide further 
possibilities to probe the Galactic potential. The SDSS presents an excellent 
data base to search for such cluster remnants.

\acknowledgments 
{\it Acknowledgements.}  
Funding for the creation and distribution of the SDSS Archive has been 
provided by the Alfred P. Sloan Foundation, the Participating Institutions, 
the National Aeronautics and Space Administration, the National Science 
Foundation, the U.S. Department of Energy, the Japanese Monbukagakusho, 
and the Max Planck Society. The SDSS Web site is http://www.sdss.org/. 
The SDSS is managed by the Astrophysical Research Consortium (ARC) for the 
Participating Institutions. The Participating Institutions are The University 
of Chicago, Fermilab, the Institute for Advanced Study, the Japan 
Participation Group, The Johns Hopkins University, Los Alamos National 
Laboratory, the Max-Planck-Institute for Astronomy (MPIA), the 
Max-Planck-Institute for Astrophysics (MPA), New Mexico State University,
University of Pittsburgh, Princeton University, the United States Naval 
Observatory, and the University of Washington.\\
M.O.\ thanks Andi Burkert for fruitful discussions and Don Schneider 
and the referee for comments that helped to improve the manuscript.

\appendix
\section{Isochrone epicycle approximation for the logarithmic potential}
Following Dehnen (1999) the isochrone approximation for the motion of a 
particle in a spherically symmetric potential $\Phi$ is obtained by 
transforming from radius $R$ and time parameter $t$ to a new radial 
coordinate $x = \sqrt{R^2 + b^2}$ and a new parameter $\eta$ with 
$dt/{d\eta} = x$. Hereby, the equation of motion changes from
\begin{eqnarray}
\frac{d^2 R}{dt^2} &=& \frac{d}{dR}\,Y  \qquad\mbox{for}\quad 
Y :=  \left(E - \Phi (R)\right) - \frac{L^2}{2R^2} 
\end{eqnarray}
to
\begin{eqnarray}
\frac{d^2 x}{d\eta^2} &=& \frac{d}{dx}\,\tilde Y\qquad\mbox{for}\quad
\tilde Y := R^2 Y \quad .
\end{eqnarray}
The quantity $\tilde Y$ as a function of $x$ is expanded into a Taylor 
series about its maximum using the first, second, and third derivative. 
For the first derivative we have
\begin{eqnarray}
\frac{d\tilde Y}{dx} &=& x \left(2\left(E-\Phi(R)\right) - 
R \frac{d\Phi}{dR}\right) \quad .
\end{eqnarray}
Since the radius $R_E$ of a circular orbit with energy $E$ is 
defined by the equation
\begin{eqnarray}
2\left(E-\Phi(R_E)\right) &=& \frac{L^2}{R_E^2} = 
R_E \left(\frac{d\Phi}{dR}\right)_{R_E} 
\end{eqnarray}
it is evident that the maximum of $\tilde Y$ lies at the value of 
$x$ that corresponds to $R_E$, i.e.,  
\begin{eqnarray}
\frac{d\tilde Y}{dx} &=& 0 \qquad\mbox{at}\quad x=x_E=\sqrt{R_E^2+b^2} .
\end{eqnarray}
Here, we focus on the special case of the logarithmic potential 
$\Phi(R) = v_c^2 \ln (R/R_0)$. Eqn.\ (A3) then reads
\begin{eqnarray}
\frac{d\tilde Y}{dx} &=& x \left(2\left(E-v_c^2 \ln (R/R_0)\right) - 
v_c^2\right) \quad .
\end{eqnarray}
The second and third derivative of $\tilde Y$ are then
\begin{eqnarray}
\frac{d^2\tilde Y}{dx^2} &=& \left(2\left(E-v_c^2 \ln (R/R_0)\right) - 
v_c^2\right) - 2 v_c^2 \frac{x^2}{R^2} \\
\frac{d^3\tilde Y}{dx^3} &=& -2 v_c^2 \frac{x}{R^2}\left(3 - 
2\frac{x^2}{R^2}\right) \quad .
\end{eqnarray}
The parameter $b$ can be chosen such that the third derivative vanishes 
in $x=x_E$. This requires  
\begin{eqnarray}
b^2 = \frac{1}{2} R_E^2 .
\end{eqnarray}
With this choice the 3rd order in the Taylor expansion of $\tilde Y$ about
$x_E$ vanishes and the error introduced by truncating 
after the quadratic term is only of order 4. The location of the 
maximum of $\tilde Y$ then is 
\begin{eqnarray}
x_E &=& \sqrt{3/2}\,R_E
\end{eqnarray}
and the second derivative 
of $\tilde Y$ at $x = x_E$ reads
\begin{eqnarray}
\left(\frac{d^2\tilde Y}{dx^2}\right)_{x_E} &=& -2 v_c^2 
\left( 1 + \frac{b^2}{R_E^2} \right) = -3 v_c^2 \quad . 
\end{eqnarray}
The expansion of $\tilde Y$ (up to third order) yields the approximate 
equation of motion
\begin{eqnarray}
\frac{d^2 x}{d\eta^2} &=& \frac{d}{dx}\left( 
-\frac{3}{2}v_c^2 \left( x - x_E \right)^2\right),
\end{eqnarray}
which is solved by a harmonic oscillation  
\begin{eqnarray}
x(\eta) &=& x_E \left(1 + e \cos \left(\sqrt{3}\,v_c\,\eta\right)\right) \quad . 
\end{eqnarray}
Here, the zero point of the parameter $\eta$ is (without loss of generality)
chosen such that it coincides with the apocenter $x=x_{max}$ 
For simplicity, we absorb the factor $\sqrt{3}\,v_c$ by setting 
$\tilde\eta := \eta \sqrt{3}\,v_c$ 
Integrating $dt/d\tilde\eta = x/(\sqrt{3}\,v_c)$ from $\tilde\eta=0$ 
to $\tilde\eta = 2\pi$ one finds the period of the oscillation to be
\begin{eqnarray}
T_R &=& \sqrt{2}\,\pi \frac{R_E}{v_c} 
\end{eqnarray}
Evaluation of $\dot{R}$ at $R=R_E$ yields 
$\dot{R}\,(R=R_E) = \pm 3 v_c e / \sqrt{2}$ and, by combination with 
the equation of energy conservation, provides the expression 
\begin{eqnarray}
e &=& \frac{\sqrt{2}}{3} \sqrt{1-\frac{L^2}{v_c^2 R_E^2}} \quad . 
\end{eqnarray}
for the eccentricity parameter $e$. By integration of 
$\dot{\varphi} = L/R^2$ the azimuth angle $\varphi$ is
\begin{eqnarray}
\varphi(\tilde\eta_2) &=& 
\sqrt{\frac{2}{3}}\frac{L}{R_E}\int_{\tilde\eta_1}^{\tilde\eta_2} 
\frac{\left(1+e\cos\tilde\eta \right)}{\left(1+e\cos\tilde\eta 
\right)^2-\frac{1}{3}} d\tilde\eta + \varphi(\tilde\eta_1) \quad .
\end{eqnarray}
If radius $R$ and absolute velocity $v$ are given for an arbitrary instant 
$t_1$ the equation of conservation of orbital energy in the logarithmic 
potential yields the parameter $R_E$ as
\begin{eqnarray}
R_E &=& R(t_1) \exp \left(\frac{1}{2}\left(
\frac{v(t_1)^2}{v_c^2} - 1 \right) \right) \quad .
\end{eqnarray}
Using this and knowing also the angular momentum $L = R(t_1) v_{\perp}(t_1)$ 
the parameter $e$ can be obtained from Eqn.\ (A15). The value of $\tilde\eta$ 
that corresponds to $t_1$ then follows from Eq.\ (A13) as 
\begin{eqnarray}
\tilde\eta_1 = \arccos \left(\frac{1}{e} \left(\sqrt{\frac{2}{3}
\frac{R(t_1)^2}{R_E^2} +\frac{1}{3}} - 1 \right)\right) \quad .
\end{eqnarray}
Finally, by integration of  $dt/d\tilde\eta = x/(\sqrt{3}\,v_c)$ the time 
$t$ for arbitrary $\tilde\eta$ is 
\begin{eqnarray}
t &=& \frac{x_E}{\sqrt{3}\,v_c} \Big[\tilde\eta + e \sin \tilde\eta 
\Big]_{\tilde\eta_1}^{\tilde\eta} + t_1 \quad .
\end{eqnarray}
In this way the radial component of the orbit and the time parameter $t$ 
are completely and explicitly determined as functions of $\tilde\eta$ for 
any given set of initial conditions. \\

\clearpage

\begin{figure*}[t]
\includegraphics[scale=0.77,bb=10 175 610 590,clip=true]{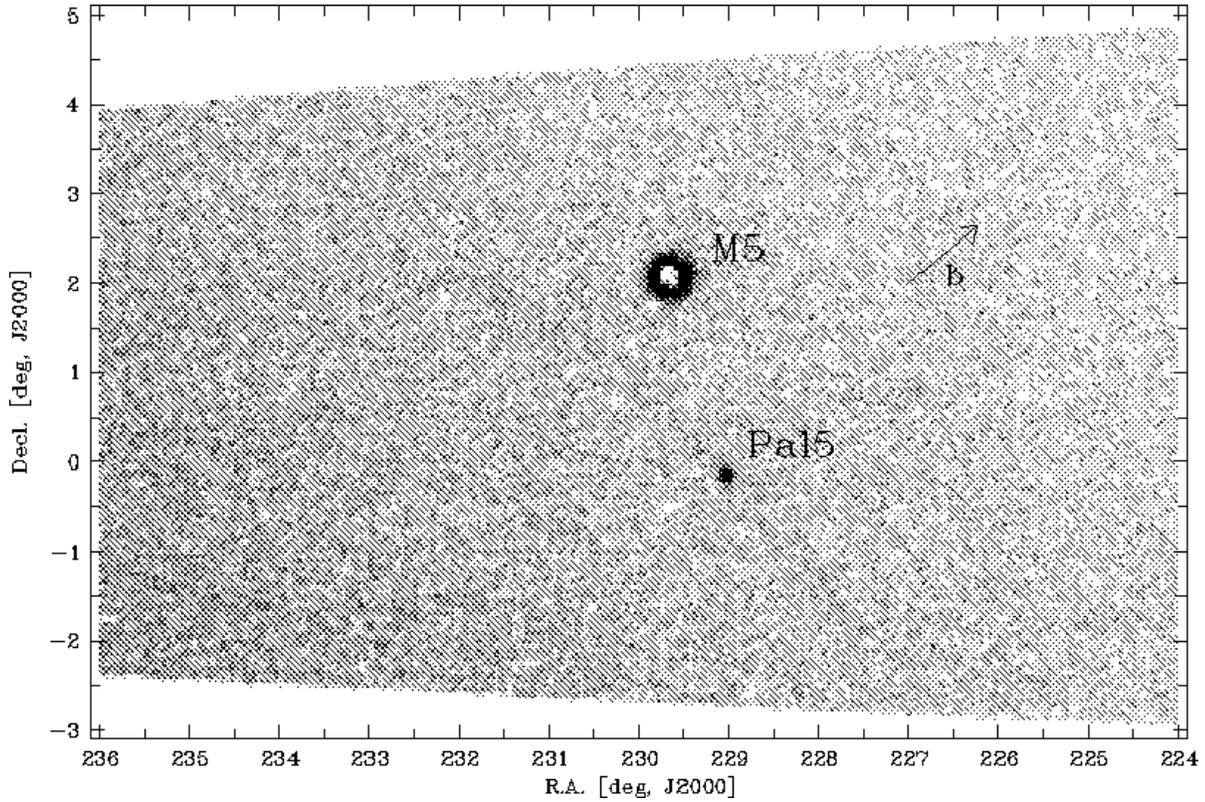}
\figcaption[Odenkirchen.fig1.ps]{Map of the surface density of SDSS point 
sources with $i^*\le 21.8$\,mag in the region of Pal\,5 (plotted vs. right 
ascension, declination). 
The density peak at position (229\fdg0,$-$0\fdg1) shows the cluster Pal\,5 
while the ring around position (229\fdg6,+2\fdg1) is due to the cluster 
M\,5 (central part incomplete because of strong crowding). Weak traces of 
tidal debris from Pal\,5 can be recognized northeast and southwest of the 
cluster. The arrow labeled with $b$ indicates the direction of increasing 
galactic latitude. For further details see \S3.1. 
\label{fig1}}
\end{figure*}

\begin{figure*}[t]
\includegraphics[scale=0.9,bb=40 460 560 678,clip=true]{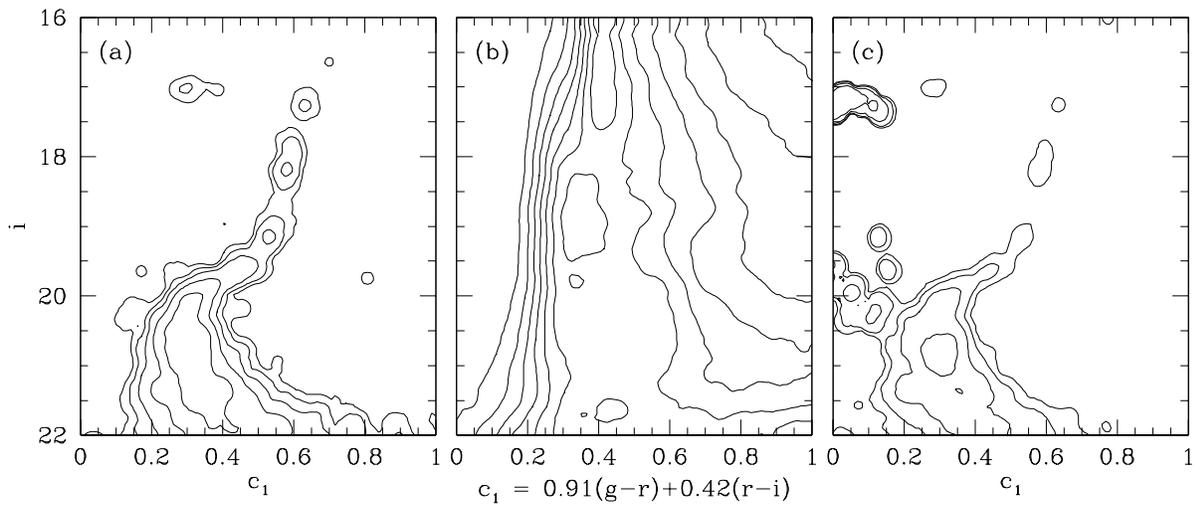}
\figcaption[Odenkirchen.fig2.ps]{Hess diagrams showing the normalized 
densities of the stellar population of Pal\,5 (a) and of the field stars 
around Pal\,5 (b) in the plane of color index $c_1$ and magnitude $i^*$, 
and the ratio of cluster to field as a function of color and magnitude (c). 
In panels (a) and (c) the contour levels increase by factors of 2 from one 
contour to the next. Panel (b) shows contours on equidistant levels. 
\label{fig2}}
\end{figure*}

\clearpage

\begin{figure*}[t]
\includegraphics[scale=0.77,bb=10 175 610 590,clip=true]{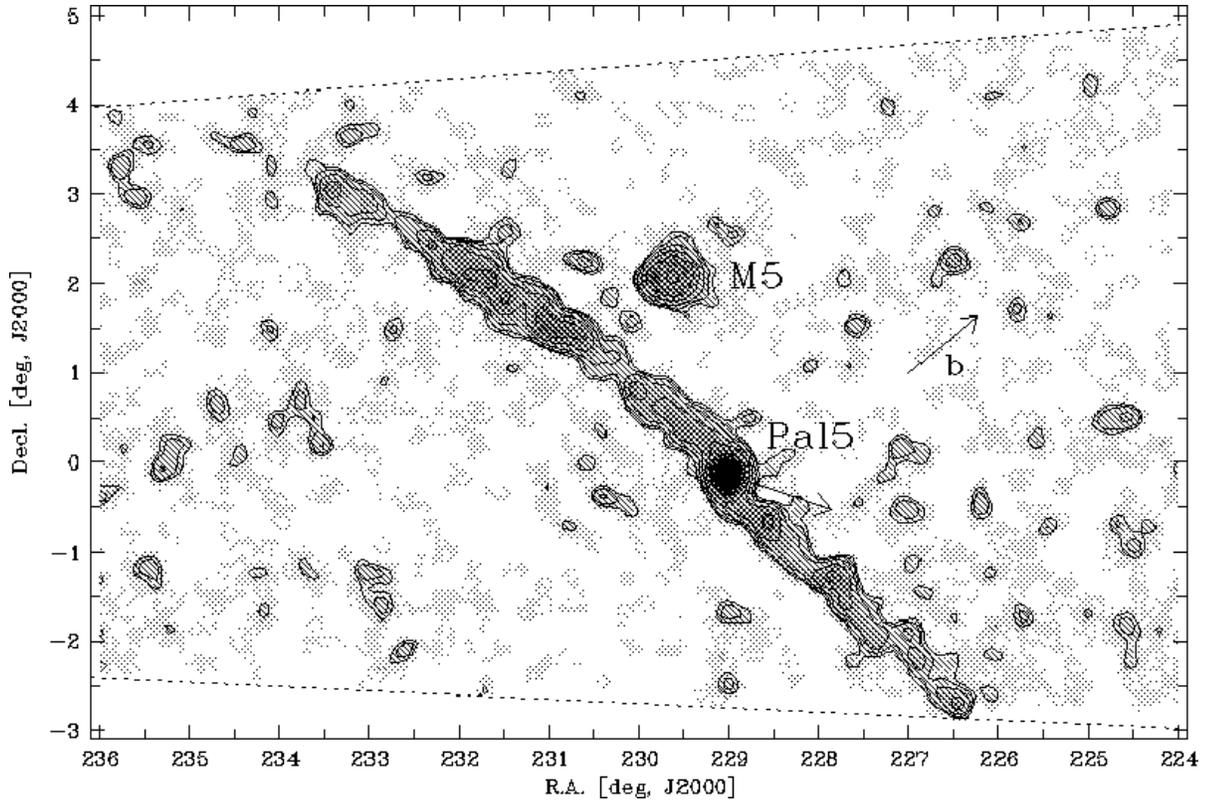}
\figcaption[Odenkirchen.fig3.ps]{Map of the surface density of stars that are 
photometrically concordant with the stellar population of Pal\,5 (plotted in 
equatorial coordinates ra,dec). 
These surface densities were derived by least-squares estimation as described 
in \S3.2. The lowest contours show density levels of $1.5\sigma$, $2\sigma$, 
$3\sigma$, and $5\sigma$ above zero (white).  
Pal\,5 is seen to be accompanied by two long tidal tails. The tidal debris 
covers an arc of almost 10$^\circ$ (for further details see \S4.1).
The arrow attached to Pal\,5 gives an approximate indication of the direction 
of its galactic motion based on the proper motion measurement by Cudworth 
(see \S5.2). The arrow labeled with $b$ shows the direction of increasing 
galactic latitude. The patch of enhanced density around (229\fdg6,+2\fdg1) 
is a residual feature from the cluster M\,5 and hence not related to Pal\,5. 
The dotted lines mark the borders of the field. 
\label{fig3}}
\end{figure*}

\begin{figure*}[t]
\includegraphics[scale=0.9,bb=30 190 480 680,clip=true]{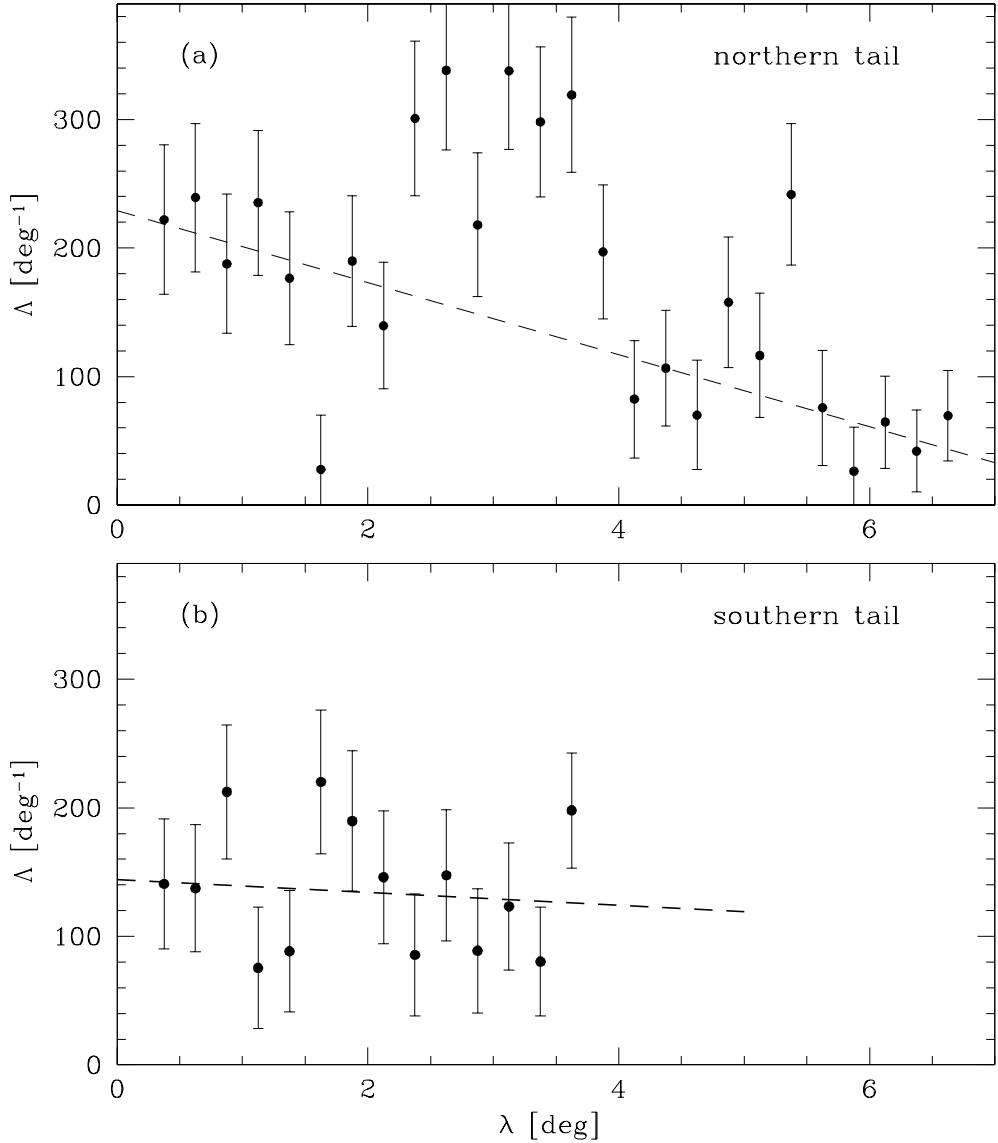}
\figcaption[Odenkirchen.fig4.ps]{Linear density $\Lambda$ of stars from 
Pal\,5 along the tidal tails of the cluster, obtained through perpendicular 
projection onto the central line of each tail. 
The parameter $\lambda$ measures the arc length along the central line of 
each tail, starting from the (projected) position of the cluster center. 
The density values are field-star subtracted. The error bars indicate the 
statistical uncertainty of the data points. The dashed lines mark the 
large-scale trend in the density derived by weighted least-squares fits.
In panel (a) the line is a fit to the five innermost and the five outermost 
data points, in panel (b) the line is a fit to all data points.  
\label{fig4}}
\end{figure*}

\clearpage

\begin{figure*}[t]
\includegraphics[scale=0.9,bb=28 266 482 610,clip=true]{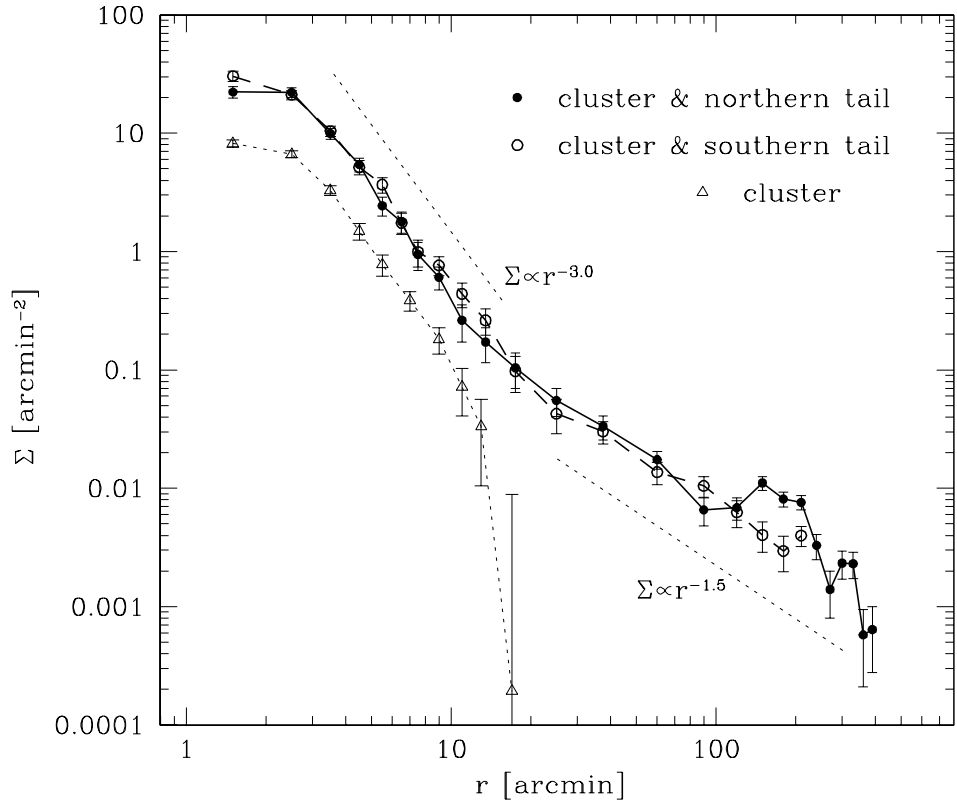}
\figcaption[Odenkirchen.fig5.ps]{Radial profile of the surface density 
$\Sigma$ of stars in Pal\,5 and its two tails (i.e., azimuthally averaged 
surface densities) from weighted number counts in annuli and annular sectors 
centered on the cluster (for details see text). For comparison, 
the open triangles show the radial density profile in two cones 
at position angles $100^\circ$ and $280^\circ$ where the contribution 
by extratidal stars is negligible (data points shifted by $-$1 
in $\log\Sigma$). 
The dashed straight lines indicate the slope of power laws with 
exponents $-$3.0 and $-1.5$. 
\label{fig5}}
\end{figure*}

\begin{figure*}[t]
\includegraphics[scale=0.9,bb=20 390 480 640,clip=true]{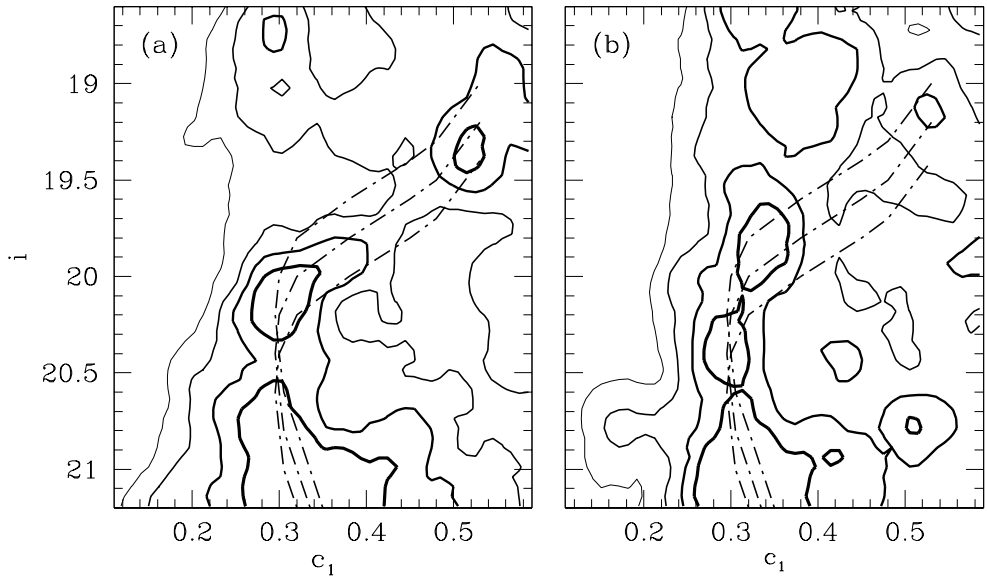}
\figcaption[Odenkirchen.fig6.ps]{Hess diagrams showing the vicinity of the 
main-sequence turn-off in two different parts of the tidal tails. 
(a) Color-magnitude distribution of stars in an $18'$ wide band (= FWHM of 
the tails) covering the northern (trailing) tail between 3\fdg5 and 5\fdg6 
from the cluster. (b) Same as (a), but for the southern (leading) tail 
between 1\fdg5 and 3\fdg6 from the cluster. The solid lines show isodensity 
contours (300, 600, 900, and 1200 stars/mag$^2$, increasing with the 
thickness of the lines). 
The dot-dashed lines mark the cluster's main-sequence and 
sub-giant branch (derived from Fig.2a), including shifts of $-$0.2, 0.0, 
and +0.2 mag in $i^*$. It is seen that the stars in the northern (trailing) 
tail lie on the same sequence as the cluster, while the stars in the outer 
part of the southern (leading) tail may on average be about 0.1 mag brighter 
in apparent magnitude. \label{fig6}}
\end{figure*}

\clearpage

\begin{figure*}[t]
\includegraphics[scale=0.9,bb=20 320 490 680,clip=true]{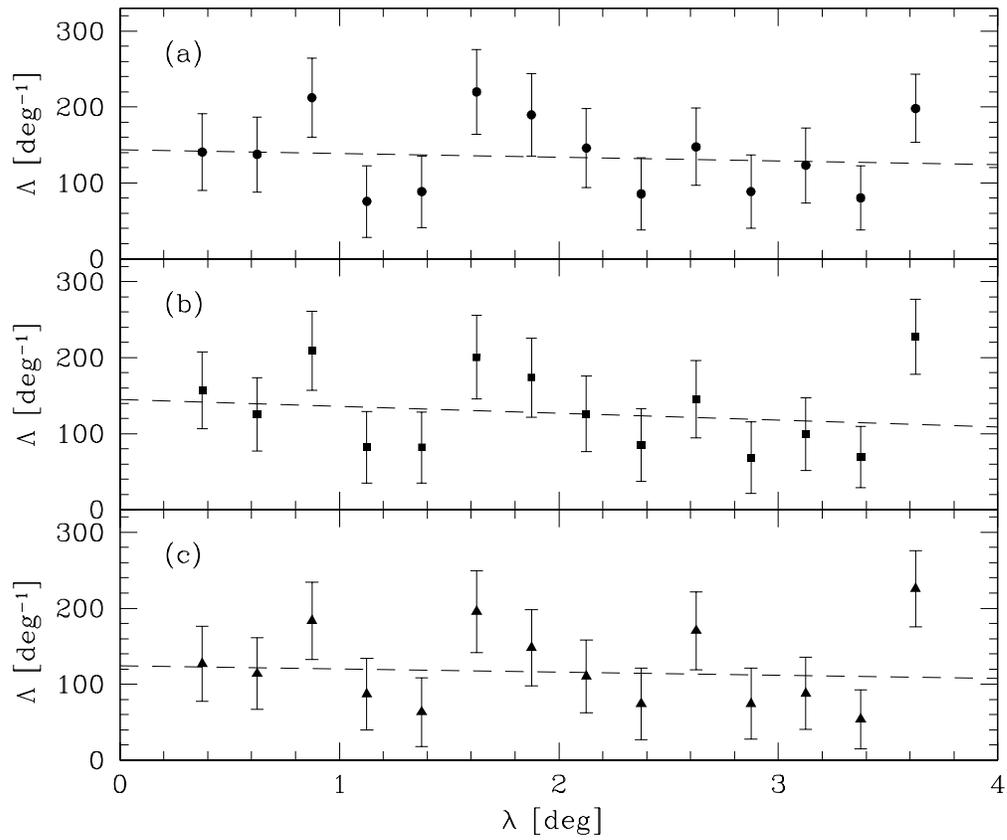}
\figcaption[Odenkirchen.fig7.ps]{Linear number density $\Lambda$ along the 
southern (leading) tail as shown in Fig.4b, 
but with certain magnitude shifts applied to the cluster color-magnitude 
template (see final paragraph of \S4.4). (a): no shift (b): after shift of 
$-$0.1~mag.(c): after shift of $-$0.2~mag. In each case the dashed line 
shows the best-fit straight line through the data points.
\label{fig7}}
\end{figure*}

\begin{figure*}[t]
\includegraphics[scale=0.8,bb=30 210 500 610,clip=true]{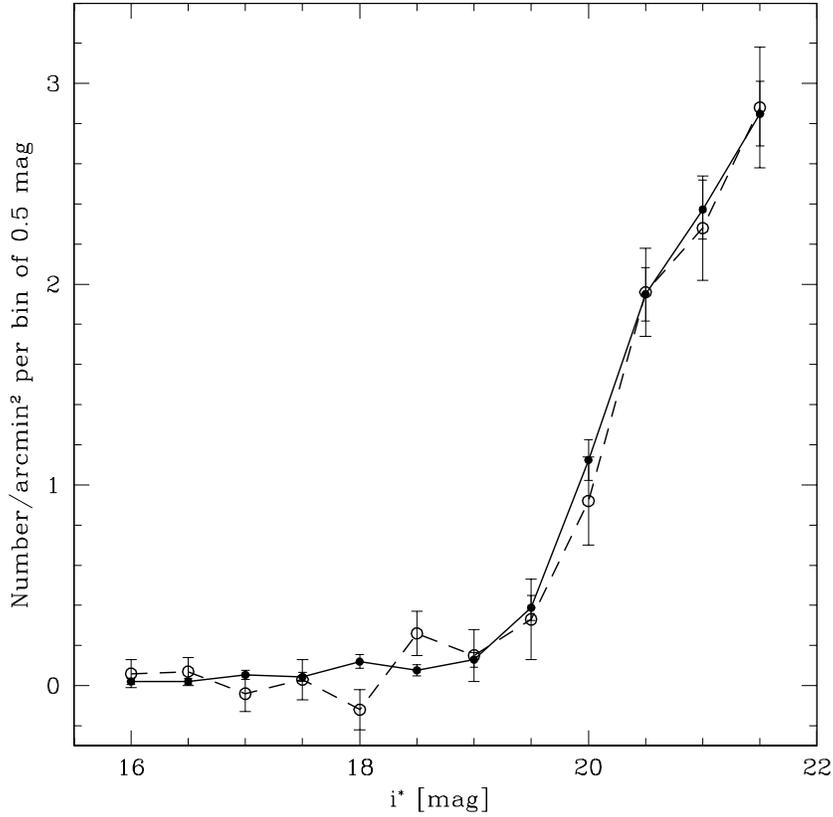}
\figcaption[Odenkirchen.fig8.ps]{Luminosity function (LF) of Pal\,5 and its 
tidal tails from star counts in a color-magnitude window comprising 
the cluster's giant branch, subgiant branch, and upper main-sequence. 
Dots/solid line: LF of stars within $r \le 6'$ from the cluster center. 
Open circles/dashed line: LF of stars in the zone of the tidal tails, 
rescaled by a factor 100 to match the LF of the cluster in the range 
$18.75\le i^* \le 19.75$. In both cases a statistical correction for 
field stars was applied.
The error bars show the statistical uncertainties of the number 
counts including the uncertainty from field star subtraction. 
Conversion to absolute magnitude can be done by $M_{i^*} = i^* - 16.8$.
\label{fig8}}
\end{figure*}

\clearpage

\begin{figure*}[t]
\includegraphics[scale=0.75,bb=90 120 430 720,angle=270,
clip=true]{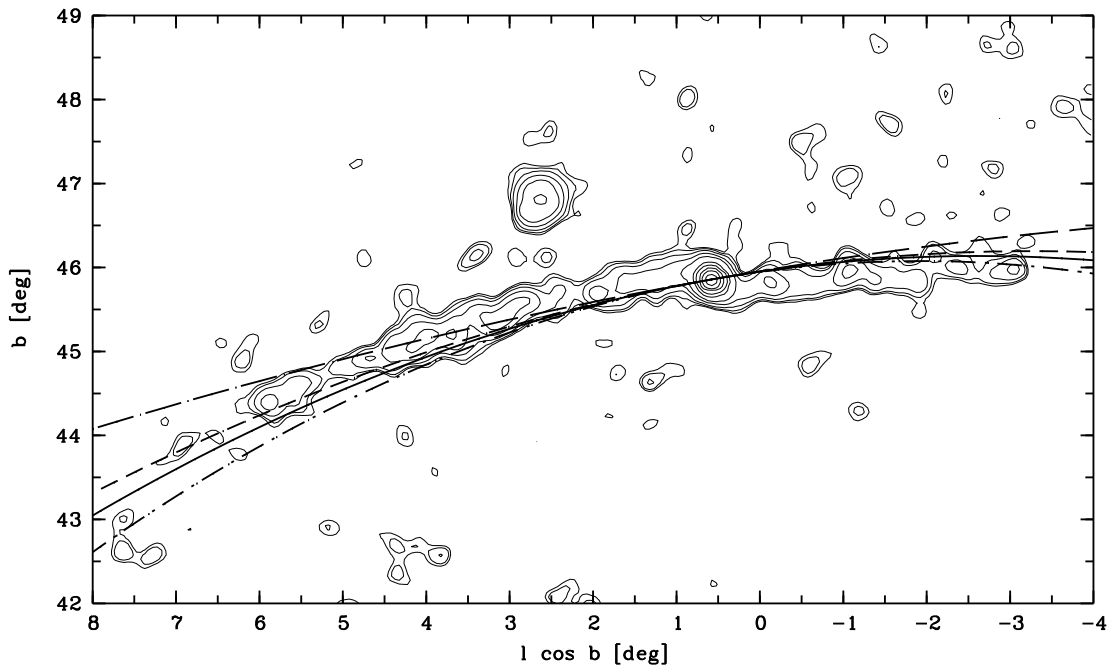}
\figcaption[Odenkirchen.fig9.ps]{The tails and the local Galactic orbit of 
Pal\,5 plotted in galactic coordinates $l \cos b, b$. 
Projections of four different orbits, all with tangent towards position 
angle $280^\circ$ at the center of the cluster, are overplotted on the 
contour map of Fig.~3. 
Long-dashed line: Straight line (i.e., unaccelerated) motion.  
Solid line: Locally best-fitting orbit in a radial field of constant 
acceleration $a = (220\,\kms)^2/18.5\,\mathrm{kpc}$.
Here, the cluster has a tangential velocity of $v_t = 95$~\kms\ 
(galactic rest frame, but viewed from the position of the Sun).
Dashed and dashed-dotted lines: Orbits in the same field, but with 
$v_t = 110$~\kms\ and $v_t = 80$~\kms, respectively.
Note that a logarithmic potential with circular velocity $v_c = 220$~\kms\ 
instead of the $a=const$ field yields projected local orbits that are  
practically identical to those shown above.
\label{fig9}}
\end{figure*}

\begin{figure*}[t]
\includegraphics[scale=0.9,bb=30 280 480 610,clip=true]{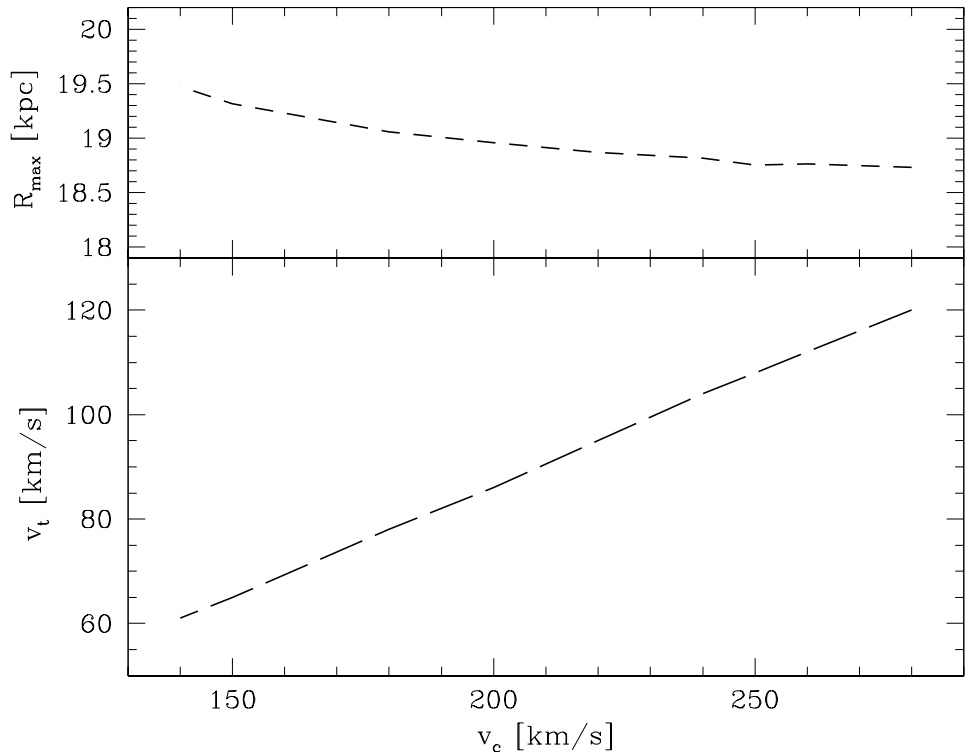}
\figcaption[Odenkirchen.fig10.ps]{Relations between the circular velocity 
$v_c$ of a spherical logarithmic potential and the present tangential 
velocity $v_t$ of the cluster Pal\,5 as well as the apocentric distance 
$R_{max}$ of the resulting orbit. These relations are obtained by choosing 
the orbit that best fits the observed tidal tails (see \S 5.2 and Fig.9). 
The increase of $v_t$ as a function of $v_c$ is linear with a slope of 0.43. 
In order to determine $v_c$ to 5 \kms\ one would need to measure $v_t$ with 
an accuracy of about 2 \kms.  
\label{fig10}}
\end{figure*}

\clearpage
 
\begin{figure*}[t]
\includegraphics[scale=0.7,bb=0 240 580 580,clip=true]{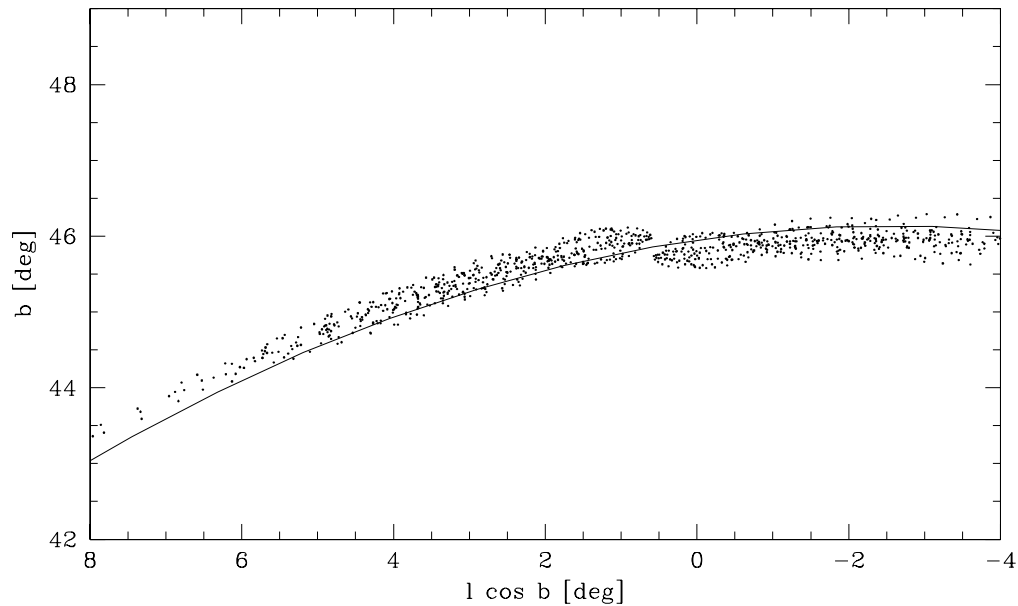}
\figcaption[Odenkirchen.fig11.ps]{Snapshot of a simulated sample of test 
particles moving in a spherical logarithmic halo potential. 
The particles were released along the orbit of Pal\,5 at equidistant 
time steps of 20 Myrs over the time-interval [$-$2;0] Gyr, with appropriate 
radial offsets in position and small radial as well as non-radial velocity 
offsets from the cluster (for details see \S 5.3). 
The dots show the positions of the particles on the sky at $t=0$ as seen from 
the Sun in galactic coordinates $l\cos b,b$. The line shows the local orbital 
path of the cluster (same as solid line in Fig.~9). It is seen that the 
stream of particles is located parallel to the orbit.
\label{fig11}}
\end{figure*}

\begin{figure*}[t]
\includegraphics[scale=0.9,bb=50 275 570 680,clip=true]{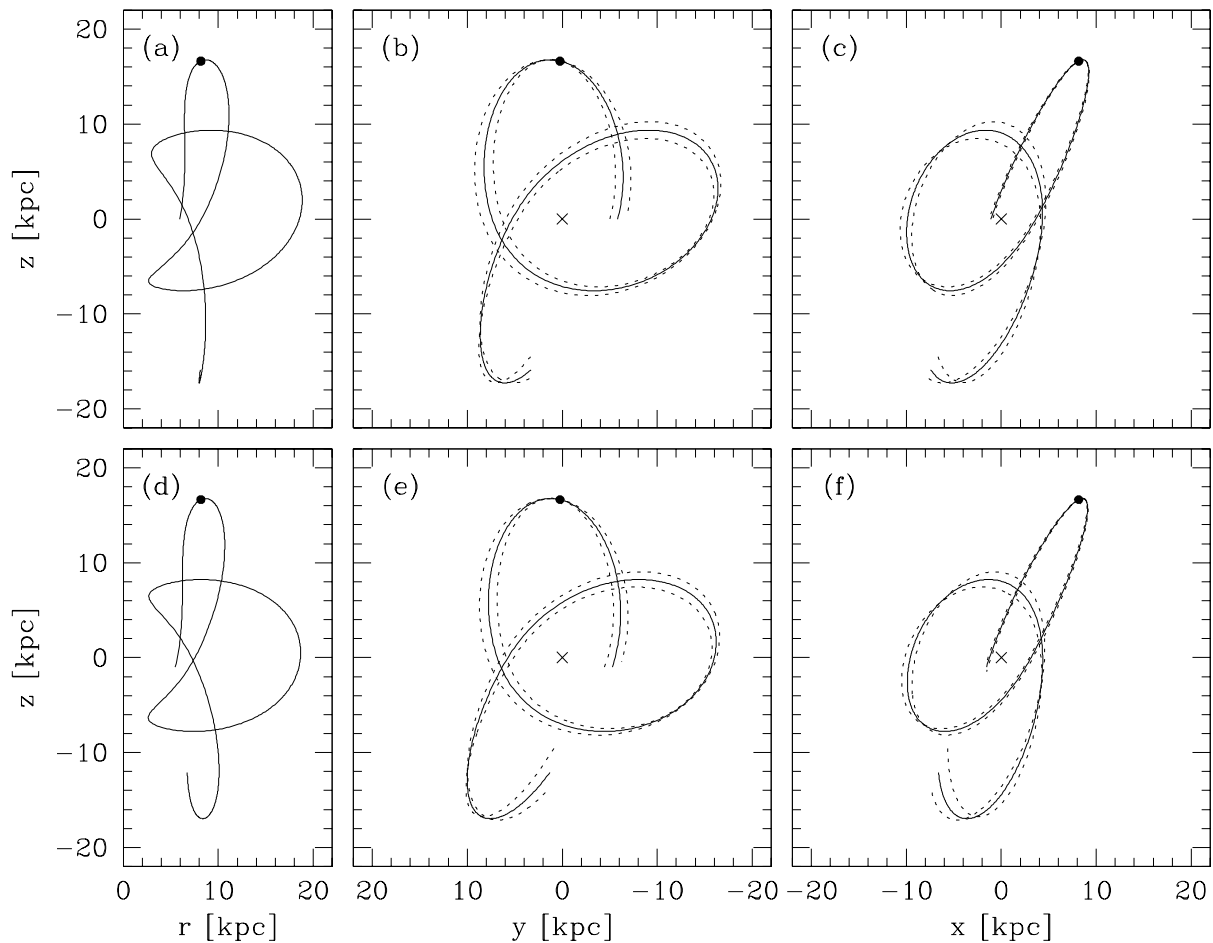}
\figcaption[Odenkirchen.fig12.ps]{The orbit of Pal\,5 in the time interval 
from $-$0.65~Gyr to +0.11~Gyr, extrapolated from the local orbit using two 
different mass models of the Milky Way. $x,y,z$ denote right-handed cartesian 
galactocentric space coordinates, with $y$ being parallel to galactic 
rotation at the Sun and $z$ pointing towards the northern Galactic pole.
Upper panels: Orbit in the Galactic potential from Dehnen \& Binney 
(1998a; Model 2).
Lower panels: Orbit in the Galactic potential from Allen \& Santillan (1991).
Left: meridional plane ($r := \sqrt{x^2+y^2}$). Middle: projection onto 
$y$-$z$ plane. Right: projection onto $x$-$z$ plane.
The present position of the cluster is indicated by the black dot. The cross 
marks the Galactic center. The solid line in all panels is for 
$v_t = 90$\,\kms, the dotted lines in the middle and right panels also show 
orbits for slightly different tangential velocities of $v_t = 80$\,\kms\ and 
$v_t = 100$\,\kms.
\label{fig12}}
\end{figure*}

\clearpage

\begin{figure*}[t]
\includegraphics[scale=0.9,bb=80 282 470 598,clip=true]{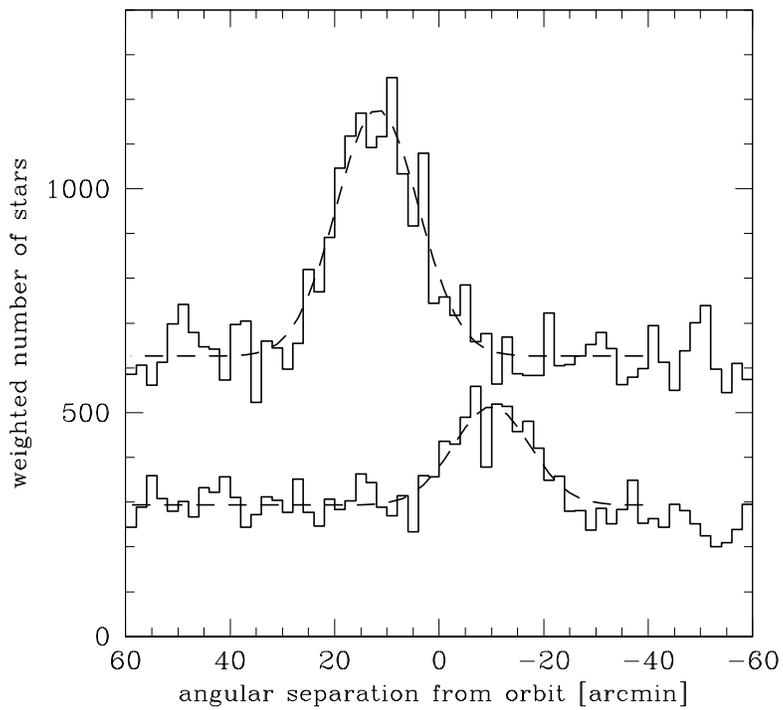}
\figcaption[Odenkirchen.fig13.ps]{Transverse offset between the tails and the 
orbit of Pal\,5 in the plane of the sky. The histograms show the weighted 
number of stars in distance bins perpendicular to the solid line of 
Figure~9. The upper histogram is for the trailing tail, the lower one for 
the leading tail. The dashed lines are best-fit Gaussians used to determine 
the centroids and widths of the distributions.  
\label{fig13}}
\end{figure*}

\begin{figure*}[t]
\includegraphics[scale=0.77,bb=10 175 610 590,clip=true]{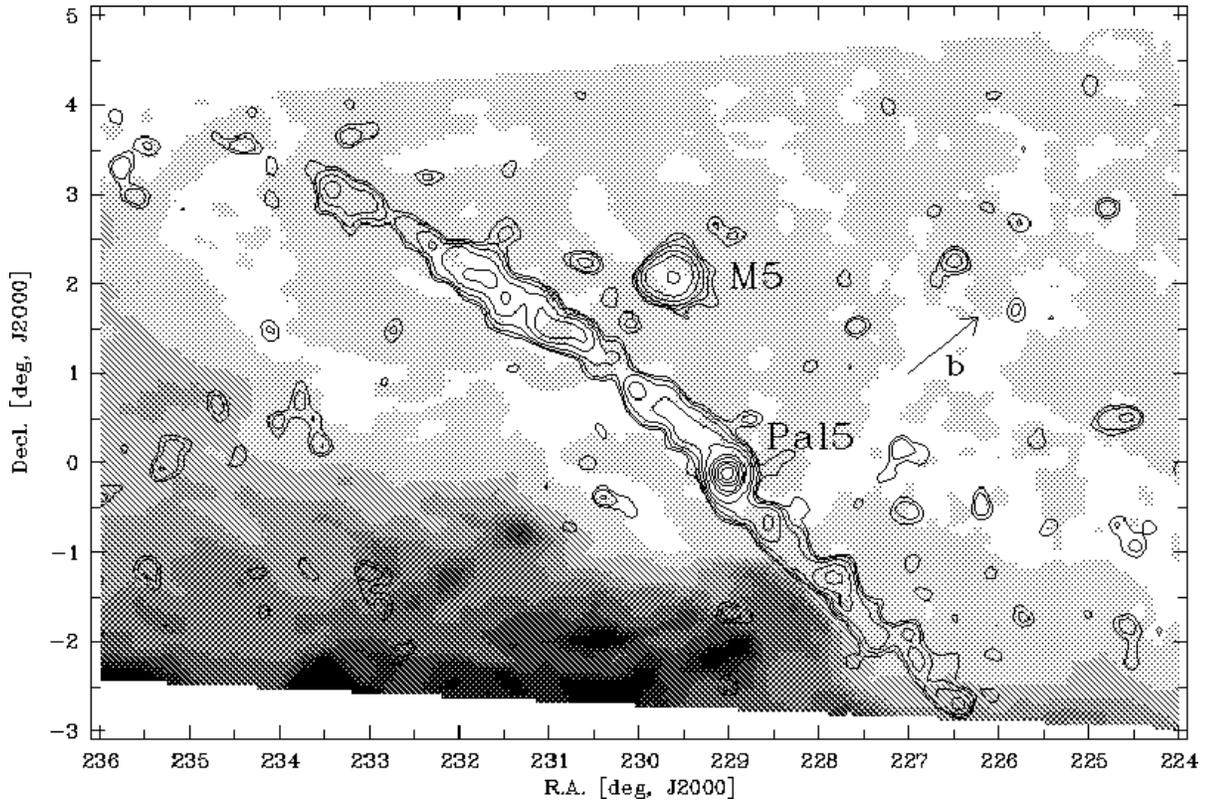}
\figcaption[Odenkirchen.fig14.ps]{Grey-scale map of interstellar extinction 
in the $g$ band based on the reddening data of Schlegel, Finkbeiner 
\& Davis (1998). The different grey scales represent extinction values 
$A_g$ from 0.15 mag (lightest grey) to 0.75 mag (black). Overlaid are the 
contours of the surface density of Pal\,5 stars shown in Figure~3. The 
arrow labeled with $b$ indicates the direction of increasing 
galactic latitude.
\label{fig14}}
\end{figure*}

\end{document}